\documentclass[twocolumn]{aastex62}

\graphicspath{{./}{figures/}}

\received{January 1, 2020}
\revised{January 7, 2020}
\accepted{\today}
\submitjournal{ApJ}

\shorttitle{M\,54's age-metallicity distribution from integrated spectroscopy}
\shortauthors{Boecker et al.}

\begin{document}

\title{Recovering age-metallicity distributions from integrated spectra: validation with MUSE data of a nearby nuclear star cluster}

\correspondingauthor{Alina Boecker}
\email{boecker@mpia.de}

\author{Alina Boecker}
\affil{Max-Planck-Institut f\"ur Astronomie, K\"onigstuhl 17, 69117 Heidelberg, Germany}

\author{Mayte Alfaro-Cuello}
\affiliation{Max-Planck-Institut f\"ur Astronomie, K\"onigstuhl 17, 69117 Heidelberg, Germany}

\author{Nadine Neumayer}
\affiliation{Max-Planck-Institut f\"ur Astronomie, K\"onigstuhl 17, 69117 Heidelberg, Germany}

\author{Ignacio Mart\'in-Navarro}
\affiliation{Max-Planck-Institut f\"ur Astronomie, K\"onigstuhl 17, 69117 Heidelberg, Germany}
\affiliation{Instituto de Astrof\'isica de Canarias, E-38200 La Laguna, Tenerife, Spain}
\affiliation{Departamento de Astrof\'isica, Universidad de La Laguna, E-38205 La Laguna, Tenerife, Spain}
\affiliation{University of California Observatories, 1156 High Street, Santa Cruz, CA 95064, USA}

\author{Ryan Leaman}
\affiliation{Max-Planck-Institut f\"ur Astronomie, K\"onigstuhl 17, 69117 Heidelberg, Germany}



\begin{abstract}

Current instruments and spectral analysis programs are now able to decompose the integrated spectrum of a stellar system into \emph{distributions} of ages and metallicities. The reliability of these methods have rarely been tested on nearby systems with resolved stellar ages and metallicities. Here we derive the age-metallicity distribution of M\,54, the nucleus of the Sagittarius dwarf spheroidal galaxy, from its integrated MUSE spectrum. We find a dominant old ($8-14$ Gyr), metal-poor (-1.5 dex) and a young (1 Gyr), metal-rich ($+0.25$ dex) component - consistent with the complex stellar populations measured from individual stars in the same MUSE data set. There is excellent agreement between the (mass-weighted) average age and metallicity of the resolved and integrated analyses. Differences are only 3\% in age and 0.2 dex metallicitiy. By co-adding individual stars to create M\,54's integrated spectrum, we show that the recovered age-metallicity distribution is insensitive to the magnitude limit of the stars or the contribution of blue horizontal branch stars - even when including additional blue wavelength coverage from the WAGGS survey. However, we find that the brightest stars can induce the spurious recovery of an old ($>8$ Gyr), metal-rich (+0.25 dex) stellar population, which is otherwise not expected from our understanding of chemical enrichment in M\,54. The overall derived stellar mass-to-light ratio of M\,54 is M/L$_{\mathrm{V}}=1.46$ with a scatter of 0.22 across the field-of-view, which we attribute to the stochastic contribution of a young, metal-rich component. These findings provide strong evidence that complex stellar population distributions can be reliably recovered from integrated spectra of extragalactic systems.

\end{abstract}



\section{Introduction}

Analyzing ages and metallicities or other chemical abundances of any stellar systems give us insight into their assembly history. Galaxies and other stellar objects \citep[e.g. nuclear star clusters (NSCs), see][for a review]{neumayer20} are assembled by a combination of in-situ secular processes such as star formation, as well as ex-situ accretion of other systems. The varying relative contribution of these two processes will lead to complex stellar populations, thus detecting and quantifying their distribution in age and metallicity is crucial to understand how galaxies and other stellar systems assemble their stellar mass. \par
Deep color-magnitude diagrams (CMDs) still count as the most reliable and detailed view that we can obtain about stellar populations present in any stellar system. Spectroscopic follow-up studies of the individual stars can then also provide radial velocities and chemical abundances. Resolved CMD analysis however automatically restricts us to within the Local Group ($\lesssim 1$ Mpc), as greater distances make it impossible to resolve stars at or below the main sequence turn-off, where most of the age information lies. As an example, the most detailed studies of stellar populations in NSCs are restricted to the nearby ones in the center of the Milky Way \citep[MW:][]{do09,do13,feldmeier-krause14,feldmeier-krause15,feldmeier-krause17a,feldmeier-krause17b} and the Sagittarius dwarf spheroidal galaxy \citep[M\,54:][]{siegel07,bellazzini08,mayte,mayte2}). \par
Because integrated broad band colors are severely prone to age-metallicity-reddening degeneracies \citep[e.g.][]{worthey94,carter09}, integrated spectra present our most detailed view of unresolved stellar systems, as different absorption lines as well as the shape of the stellar continuum respond more sensitively to age and abundance pattern changes \citep{sanchez-blazquez11}. With the advancement in sensitivity, wavelength coverage and spectral resolution of spectrographs (\citealp[e.g. XShooter:][]{xshooter}) or the new generation of integral-field units (\citealp[e.g. MUSE:][]{muse}), the analysis of integrated spectra is changing from line strength analysis \cite[e.g.][]{worthey94,worthey97,thomas03,thomas05,trager00a,trager00b} to fitting the whole observed spectrum maximizing the information present in each spectral pixel \cite[e.g.][]{cidfernandes13,wilkinson15,mcdermid15,comparat17,goddard17a,chauke18,kacharov18}. \par
Even though integrated line strength analysis is still crucial in understanding certain (galaxy) features sensitive to specific spectral lines such as the initial mass function (IMF) or individual $\alpha$-element abundances \citep[e.g.][]{martin-navarro18a,martin-navarro19}, it has been shown that single stellar population (SSP) equivalent ages and metallicities are biased towards the youngest stellar population present in the integrated light \cite[e.g.][]{serra07,trager09}. With full spectral fitting methods however, we are slowly moving towards uncovering the whole chemical enrichment history of a stellar system by fitting a linear combination of multiple SSP models to the observed spectrum. Yet, the techniques necessary to not only recover mean ages and metallicities, but also a \emph{distribution} in that parameter space \cite[e.g.][]{ppxf04,starlight1,stecmap,steckmap,ppxf17,wilkinson17} are still under development. On top of that, additional factors influencing their performance, e.g. the wavelength coverage or spectral resolution are also not well understood yet. However, deriving age-metallicity distributions are crucial in deciphering the mass assembly of extragalactic systems \citep[e.g.][]{boecker19}. \par
Given these challenges, it would be beneficial to study resolved and integrated spectra of the same system to test the reliability of age-metallicity distribution recovery with full spectral fitting methods. Here, we utilize the richness of the $3.5^{\prime}\times 3.5^{\prime}$ MUSE \citep[Multi-Unit Spectroscopic Explorer:][]{muse} data set of M\,54, the nucleus in the Sagittarius dwarf spheroidal galaxy, in order to apply the full spectral fitting to its integrated spectrum aiming to recover its multiple populations and compare the results to the resolved study of the same data set \citep{mayte}. With respect to similar comparisons between resolved and integrated SFHs \citep{ruiz-lara15,ruiz-lara18}, our advantage lies in the power of the MUSE instrument \citep[see also][]{kuncarayakti16}: a) the two approaches are performed using the same data, which means that possible instrumental effects stay the same, b) the metallicity of each star can be directly derived from its individual spectrum, which are regarded intrinsically more reliable than photometric metallicities and thus reduce degeneracies further, when stellar ages are determined from isochrone fitting. \par
This paper is organized as follows: in Section \ref{sec: data} we present the three different integral field unit (IFU) data sets analyzed in this work and briefly describe the analysis of the resolved stars from \citet{mayte}; in Section \ref{sec: analysis} we describe our analysis method of deriving age-metallicity distributions from integrated spectra; in Section \ref{sec: tests1} we show the results of this technique in dependence of different integrated spectra of M\,54; in Section \ref{sec: results} we compare our integrated analysis method with the resolved star analysis; in Section \ref{sec: discussion} we discuss our results and in Section \ref{sec: conclusion} we give our conclusions of this comparison exercise.

\section{IFU Data of M\,54}\label{sec: data}

\begin{figure*}[htbp]
    \centering
    \includegraphics[width=1.5\columnwidth]{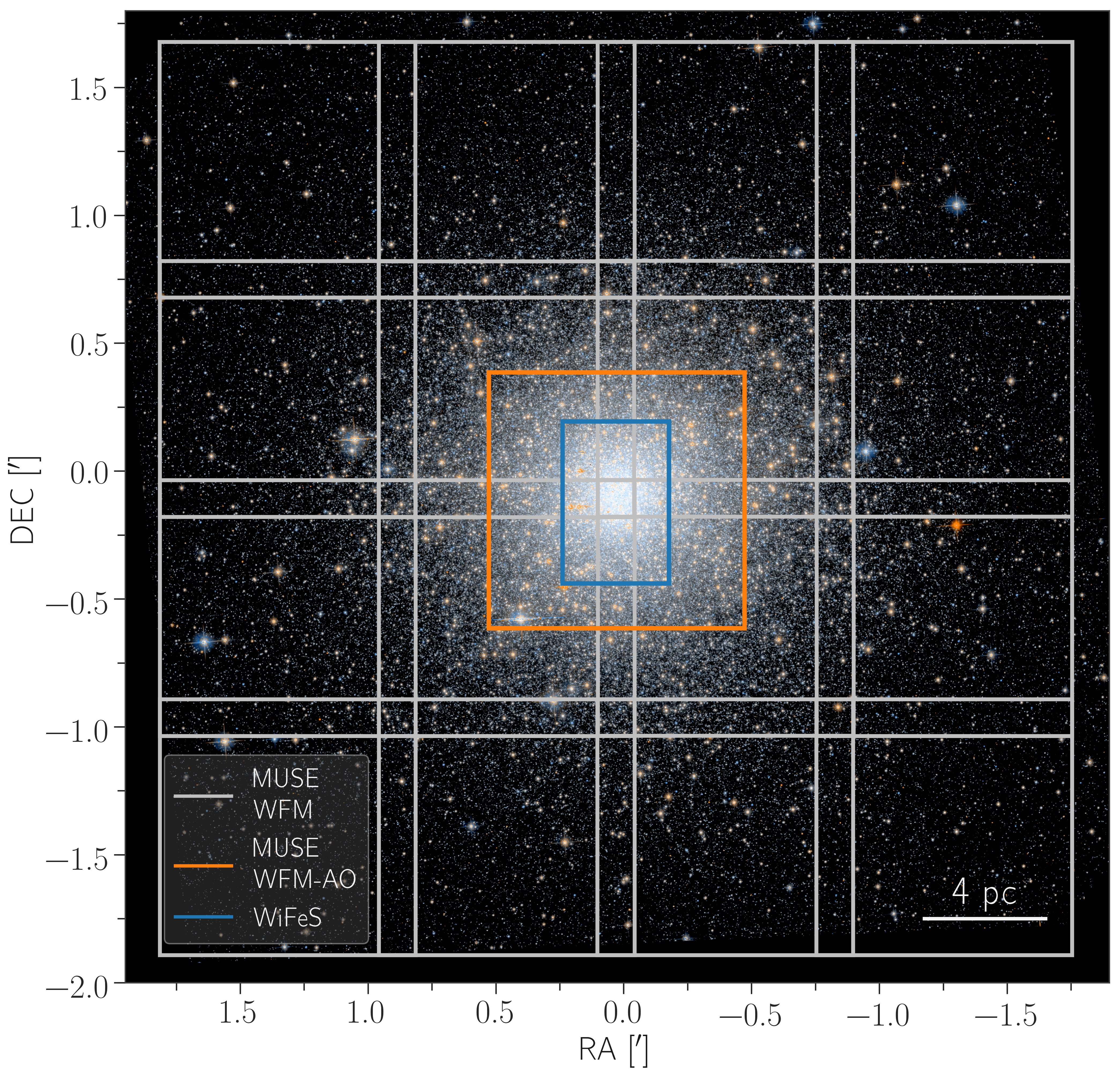}
    \caption{Color image from HST ACS/WFC in the F606W \& F814W filters \protect\citep{sarajedini07} of M\,54 overlaid with the pointings of the three integral field unit data sets used in this work: MUSE WFM (light grey), MUSE WFM-AO (orange) and WiFeS (blue).}
    \label{fig: data}
\end{figure*}

In this Section we briefly describe the data we use to analyze the stellar populations of M\,54 from its integrated spectrum. While our comparison between resolved and integrated stellar population extraction is focused on the MUSE WFM data from \citet{mayte}, we also include data from the MUSE WFM-AO science verification and the publicly available WAGGS (WiFeS Atlas of Galactic Globular cluster Spectra; \citealt{waggs1}) survey exploiting different instrument systematics like wavelength coverage and spectral resolution. In Figure \ref{fig: data}, we show the pointings on M\,54 of the three data sets as an overview.

\subsection{MUSE WFM}\label{sec: mayte_analysis}

\citet{mayte} analyzed a $4 \times 4$ MUSE mosaic centered on M\,54 (095.B-0585(A), PI: L\"utzgendorf) covering a total area of $3.5^{\prime}\times 3.5^{\prime}$, which corresponds to an extension of about 2.5 times the cluster's effective radius ($r_{eff}=0.82^{\prime}\ \hat{=}\ 6.78$ pc \citep[][2010 edition]{harris} at a distance of 28.4 kpc \citep{siegel11}). MUSE is an integral field spectrograph, mounted on the UT4 of the Very Large Telescope at the Paranal Observatory in Chile. It has a wavelength coverage of $4750-9300$ \AA \ with 1.25 \AA/pix sampling and a mean spectral resolution of $\sim 3000$. More information about the observing strategy and data reduction can be found in \citet{mayte}. \par
To derive individual ages and metallicites for the member stars of M\,54 and identify multiple populations, \citet{mayte} performed five essential steps, which are stated here in short for clarity:
\begin{enumerate}
    \item Extract individual spectra of the resolved stars with a wavelength dependent PSF-weighting technique \citep[PampelMuse:][]{kamann13} by using a photometric reference catalogue from HST \citep{siegel07}. All spectra with a $\mathrm{SNR} < 10$ are excluded from further analysis. This corresponds to a limiting magnitude of $\mathrm{I}=22$ mag, which includes stars just below the turn-off region.
    \item Fit all extracted spectra with a full spectral fitting software \citep[ULySS:][]{ulyss} and a stellar model library \citep[ELODIE 3.2:][]{elodie32} to determine atmospheric parameters ($\text{T}_{\text{eff}}$, $\log \text{g}$, [Fe/H]) and the radial velocities. This was done for the wavelength range of $4750-6800$ \AA.
    \item Determine member stars of the cluster by using an iterative expectation maximization technique \citep[clumPy:][]{kimmig15} based on the position and the radial velocity measurement of each star. Stars with a membership probability of $\geq 70\%$ are considered to belong to M\,54.
    \item Estimate individual stellar ages by fitting scaled-solar isochrones \citep[from the Dartmouth Steller Evolution Database:][]{dartmouth} using HST photometry (F606W \& F814W filters) and their spectroscopically derived iron abundances within a Bayesian framework. Horizontal branch stars were excluded from this.
    \item Perform Gaussian mixture models in the derived age-metallicity parameter space to determine the most likely number of distinct stellar populations present in M\,54 as well as the probability of each member star belonging to one of those populations.
\end{enumerate}

\subsection{MUSE WFM-AO}

MUSE WFM is also offered with ground layer adaptive optics correction by the GALACSI (Ground Atmospheric Layer Adaptive Corrector for Spectroscopic Imaging) module aimed to double the ensquared energy in one $0.2^{\prime}\times 0.2^{\prime}$ spaxel as compared to natural seeing. Due to the four laser guide stars the wavelength range around the NaD lines ($5820-5970$ \AA) is blocked. We include the analysis of MUSE WFM-AO science verification data (60.A-9181(A), PI: Alfaro-Cuello) of M\,54 investigating, whether the Na notch filter has an influence on the stellar population inference. The data consists of a single, central $1^{\prime}\times 1^{\prime}$ pointing.

\subsection{WAGGS survey}\label{sec: waggs}

MUSE does not cover blue wavelengths short of 4750 \AA, which is often raised as caveat \citep[see planned BlueMUSE:][]{bluemuse}, since it is commonly understood that young stellar populations ($< 1$ Gyr) or certain stellar evolutionary stages, such as horizontal branch stars, dominate at these bluer wavelengths ($< 4000$ \AA). It also misses other important spectral lines like the Ca H \& K lines, at 3969 \AA \ and 3934 \AA \ respectively, which are particularly sensitive to metallicity changes at fixed temperature. Therefore their potential contribution to the overall integrated light might not be significant enough in the MUSE wavelength range. To test this, we additionally analyze the integrated spectrum of M\,54 from the WAGGS survey \citep{waggs1}. \par
The goal of the survey is to provide a library of globular cluster (GC) spectra in the Milky Way and its satellite galaxies with a higher resolution ($\mathrm{R}\sim 7000$) and wider wavelength coverage ($3270-9050$ \AA) than other studies to investigate their stellar populations in detail. For this purpose, they utilize WiFeS (Wide Field Spectrograph), a dual arm integral field spectrograph, on the Australian National University 2.3m telescope at the Siding Spring Observatory, which has a field of view of $38^{\prime\prime}\times25^{\prime\prime}$, hence targeting the center of the GCs. The spectrograph offers four, high resolution gratings, U7000 ($3270-4350$ \AA\ with 0.27 \AA/pix), B7000 ($4170-5540$ \AA\ with 0.37 \AA/pix), R7000 ($5280-7020$ \AA\ with 0.44 \AA/pix) and I7000 ($6800-9050$ \AA\ with 0.57 \AA/pix) in order to achieve the large wavelength coverage. For more information about the observing strategy and data reduction see \citet{waggs1}. The integrated spectra for all their observed GCs are publicly available on their website\footnote{\url{http://www.astro.ljmu.ac.uk/~astcushe/waggs/data.html}}.\par
As the spectra from WAGGS consist of four parts corresponding to the four gratings, we determine the new flux in the overlapping regions as the error weighted mean in order to generate one continuous spectrum. The corresponding new inverse error in that region is the mean of the inverse of the two overlapping error spectra. Then we re-sample the entire spectrum to the highest pixel dispersion of 0.57 \AA/pix \citep[using SpectRes:][]{spectres}. \par

\section{Method for analyzing an integrated spectrum}\label{sec: analysis}

In this Section, we briefly describe our approach of full spectral fitting and how multiple stellar populations in age-metallicity space can be derived from an integrated spectrum. In theory, there are only two ingredients needed: a single stellar population spectral library and a fitting machinery that fits the models to the data.

\subsection{Full spectral fitting method}

We can extract stellar populations properties from an integrated spectrum by viewing the integrated light as a linear combination of many single stellar populations, each with a different, single age and metallicity\footnote{Here, we keep the IMF and heavy element abundance fixed.}. Hence, the full spectral fitting algorithm finds the optimal weight for each SSP spectrum, such that their sum best represents the observed integrated spectrum. Since SSP models are normally normalized to one solar mass, the best-fit weights are mass fractions. \par
Due to the vast parameter space of the SSPs (typically $> 500$ models) and the typical age-metallicity degeneracy, this inverse problem is usually ill-posed and ill-conditioned. In our case, ill-posed means that the solution is not necessarily unique, as many different SSP combinations can represent the data equally well. Ill-conditioned refers to the fact that fluctuations on the noise-level in the data can drastically change the solution. Regularized least-squares minimization is a common way to treat both of these issues. This technique is also implemented in the program pPXF \citep{ppxf04,ppxf17}, which we will use in this study. It has the advantage of being able to derive two dimensional age-metallicity distributions, which is not the case for other fitting algorithms that use similar approaches \citep[e.g. STECKMAP:][]{stecmap,steckmap}. \par
In the context of pPXF, regularization provides a way to smoothly link the sparsely returned unregularized mass fractions in age-metallicity space until a certain criterion on the data fidelity is met. This smoothing may be well motivated in the case of galaxies, where we assume that chemical enrichment does not likely occur in a discrete manner. However, if the solution requires a more bursty star formation history, and the data quality is high enough, regularization will allow for that as well \citep{ppxf17}. \par
How \emph{much} the weights become smeared out is controlled by the regularization parameter ($\lambda$), whereas the \emph{way} they are distributed is imposed by the regularization matrix (\textbf{B}), which is typically a finite difference operator. We note that we use the first finite difference ($\mathbf{B}=\mathrm{diag}(1,-1)$) throughout this work, however we made sure that the second and third order one give consistent results \citep[see e.g.][]{stecmap,regulmatrix,kacharov18,boecker19}. \par
In any case, the essential pPXF fitting procedure is always the same, which follows the instruction in the source code of pPXF as well as \citet[][Chapter 19.5]{press07}. First, an unregularized fit is performed in order to re-scale the noise vector, such that this fit has a reduced $\chi^2$ of unity. Then regularized fits are performed, tuning the regularization parameter such that the $\chi^2$ of the regularized fit moves one standard deviation away, which corresponds to $\sqrt{2\cdot\#\mathrm{pixel}}$. This sets a regularization parameter, which allows for a maximum amount of smoothness that is still regarded to be consistent with the data \citep[see e.g.][]{mcdermid15,kacharov18,boecker19}. \par
Before the first fit, we conduct the following, preparatory steps: SSP models are broadened to the wavelength-dependent line spread function (LSF) of MUSE\footnote{$\mathrm{FWHM}(\lambda)=5.866\times 10^{-8} \lambda^2 - 9.187\times 10^{-4} \lambda+6.040$ \citep[from][Figure 4]{guerou17}.}, as the SSP models have slightly higher spectral resolution in the blue part of the spectrum than MUSE. In case of the WAGGS spectrum, we broaden the observed spectrum to a constant FWHM of 2.5 \AA, which is the approximate average spectral resolution of MUSE. After that the wavelength range of interest is selected and residual skylines are masked out. Solely multiplicative polynomials are used to correct for any continuum mismatch between the SSP models and the observed spectrum. Their degree is determined according to $\lfloor{(\lambda_{\mathrm{max}}-\lambda_{\mathrm{min}})/200\ \text{\AA}}\rfloor$ ensuring that spectral features narrower than 200 \AA \ are not influenced by the polynomial. For every fit we also include the corresponding error spectrum as determined by the data reduction process. In pPXF, all inputs have to be logarithmically re-binned in wavelength before the fitting process. \par
Together with the mass weights of the bestfit returned by pPXF and the predictions for the total luminosity of each SSP model in a certain photometric band, we can calculate the mass-to-light ratio (M/L) of the stellar system. We follow \citet[][equation 2]{cappellari13}
\begin{equation}\label{eq: ml}
    \mathrm{M}/\mathrm{L}_{\mathrm{V}}=\frac{\sum_i w_i M^{\star+rem}_i}{\sum_i w_i L_{V,i}}
\end{equation}
where, for the $i$-th SSP model, $w$ are the weights returned by pPXF, $M^{\star+rem}$ is the mass in stars and dark remnants\footnote{The total mass of one SSP model is by construction 1 $\mathrm{M}_{\odot}$. $M^{\star+rem}$ is typically $\neq1$ $\mathrm{M}_{\odot}$ due to the mass loss during stellar evolution.} and $L_V$ the corresponding V-band luminosity. \par
We can also estimate the total stellar mass of the system from the integrated spectrum. For this we need to take into account the distance $d$ to the stellar object, as well as normalization constants applied to the observed spectrum $N_{obs}$ and the SSP models $N_{SSP}$ prior to fitting\footnote{This typically means diving the spectra by the median in order to avoid numerical artifacts.}. Hence, we arrive at the following formula \citep[see also][]{wilkinson17,kacharov18} for the total stellar mass $\mathrm{M_{\star,\ \mathrm{tot}}}$:
\begin{equation}\label{eq: mass}
    \mathrm{M_{\star,\ \mathrm{tot}}}=\sum_i M_{SSP, i}=4\pi d^2\frac{N_{obs}}{N_{SSP}} \sum_i w_i M^{\star+rem}_i
\end{equation}

\subsection{Single stellar population models}\label{sec: SSP}

Due to the large age and metallicity as well as wavelength coverage ($1680-50000$ \AA) we chose the SSP models from the E-MILES library \citep{vazdekis16} for our main analysis of M\,54's stellar populations. Using the BaSTI isochrones \citep{basti04} they cover 53 age bins between $0.03-14.0$ Gyr and 12 metallicity [M/H] bins between -2.27 and 0.4 dex. We are using the bimodal IMF with a slope of 1.3 \citet{vazdekis96,vazdekis03}, which is similar to a Kroupa \citep{kroupa01} IMF. \par
The spectral resolution of the E-MILES models is 2.51 \AA \ (FWHM) in the MUSE wavelength range until 8950.4 \AA, after that it jumps to about 4.2 \AA \ and then increases slowly with wavelength \citep[see][Figure 8]{vazdekis16}. This means that the LSF of MUSE is much narrower between $8950.4-9300$ \AA \ than the SSP models and therefore we truncate the MUSE spectrum there. Additionally, the contamination from sky residuals can be quite significant in this regime anyway. How this choice of truncation influences the age-metallicity recovery is investigated in Appendix \ref{appendix1}. \par
In Section \ref{sec: discuss1} we will discuss the impact of using different SSP models for our analysis. However, it is beyond the scope of this paper to provide a full comparison among different SSP libraries and their impact on the recovery of the age-metallicity distribution from integrated spectra. Thus, the reader is referred to more detailed studies of different spectral synthesis assumptions and techniques in the context of full spectral fitting \citep[e.g.][]{conroy09,conroy10a,conroy10b,gonzalezdelgado10,cidfernandes10,fan16,baldwin18,dahmer-hahn18,ge18,ge19}.

\section{Stellar population results for different integrated spectra of M\,54}\label{sec: tests1}

\begin{deluxetable*}{ccccc}[htb!]
    \tablecaption{Different integrated spectra of M\,54.\label{tab: spectra}}
    \tablecolumns{4}
    \tablewidth{0pt}
    \tablehead{\colhead{Set} & \multicolumn{2}{c}{Data} & \colhead{SNR per \AA} & \colhead{Figure}}
    \startdata
    A & MUSE WFM &  all 16 cubes & 148\tablenotemark{a} & \ref{fig: all_cubes} \\
    B & MUSE WFM & single cubes 1-16 & 125, 114, 119, 80, 79, 149, 141, 111, 110, 127, 148, 105, 79, 77, 94, 104\tablenotemark{a} & \ref{fig: single_cubes} \\
    C & MUSE WFM & single stars & 100 (all stars), 113 (members only)\tablenotemark{a} & \ref{fig: single_stars} \\
    D & MUSE WFM & single stars & 77 ($\mathrm{I}\leq 16$ mag), 104 ($\mathrm{I}\leq 18$ mag), 114 ($\mathrm{I}\leq 20$ mag)\tablenotemark{a} & \ref{fig: magnitude_cut} \\
    E & MUSE WFM-AO & whole cube & 134\tablenotemark{a} & \ref{fig: all_cubes} \\
    F & WAGGS & whole cube & 86, 136, 366, 199\tablenotemark{b} & \ref{fig: waggs1} \\
    \enddata
    \tablenotetext{a}{The SNR was estimated from the pPXF fit residuals between 5000 \AA \ and 5500 \AA. Due to the correlations in the noise and systematic effects from the data reduction, the signal-to-noise is heavily overestimated, when determined from the formal error cube. We therefore follow this more conservative approach as done in \protect\citet{califadr2,fornax3d}.}
    \tablenotetext{b}{The SNR for the U,R,B,I gratings respectively was taken from \protect\citet[][Table 2]{waggs1}. Although the SNR seems to be higher than from the MUSE observations, by looking at the residuals of Figure \ref{fig: waggs1} it becomes apparent that they are of the same order.}
\end{deluxetable*}

The flux-calibration of MUSE data makes it straightforward to construct an integrated spectrum of M\,54. We can either sum up the individually extracted stars or collapse the whole data cube along the spatial axes. \par
In this Section we will probe the recovery of M\,54's multiple stellar populations depending on the details of constructing the integrated spectrum from the MUSE data and on instrumental effects like wavelength coverage and spectral resolution by analyzing the WAGGS spectrum. \par
We summarize in Table \ref{tab: spectra} the different integrated spectra of M\,54 analyzed in this work.

\subsection{Integrated spectra from the entire MUSE cube}\label{sec: cubes}

Firstly, we construct an integrated spectrum from the full cube for each of the 16 MUSE WFM pointings (see data set B in Table \ref{tab: spectra}). Only spaxels with a formal $\mathrm{SNR}>10$ are included in the total integrated spectrum in order to avoid heavy sky residuals in the final integrated spectrum. Those are especially apparent for outer pointings. We do not exclude any possible contaminating sources and tested that the SNR cut does not affect the stellar populations recovery. We additionally combine these 16 integrated spectra into one single integrated spectrum (see data set A in Table \ref{tab: spectra}). Note that we do not account for the overlapping regions of the 16 individual pointings (see Figure \ref{fig: data}). The same is done for the single MUSE WFM-AO cube (see data set C in Table \ref{tab: spectra}). \par
\begin{figure*}[htbp]
\centering
\includegraphics[width=\textwidth]{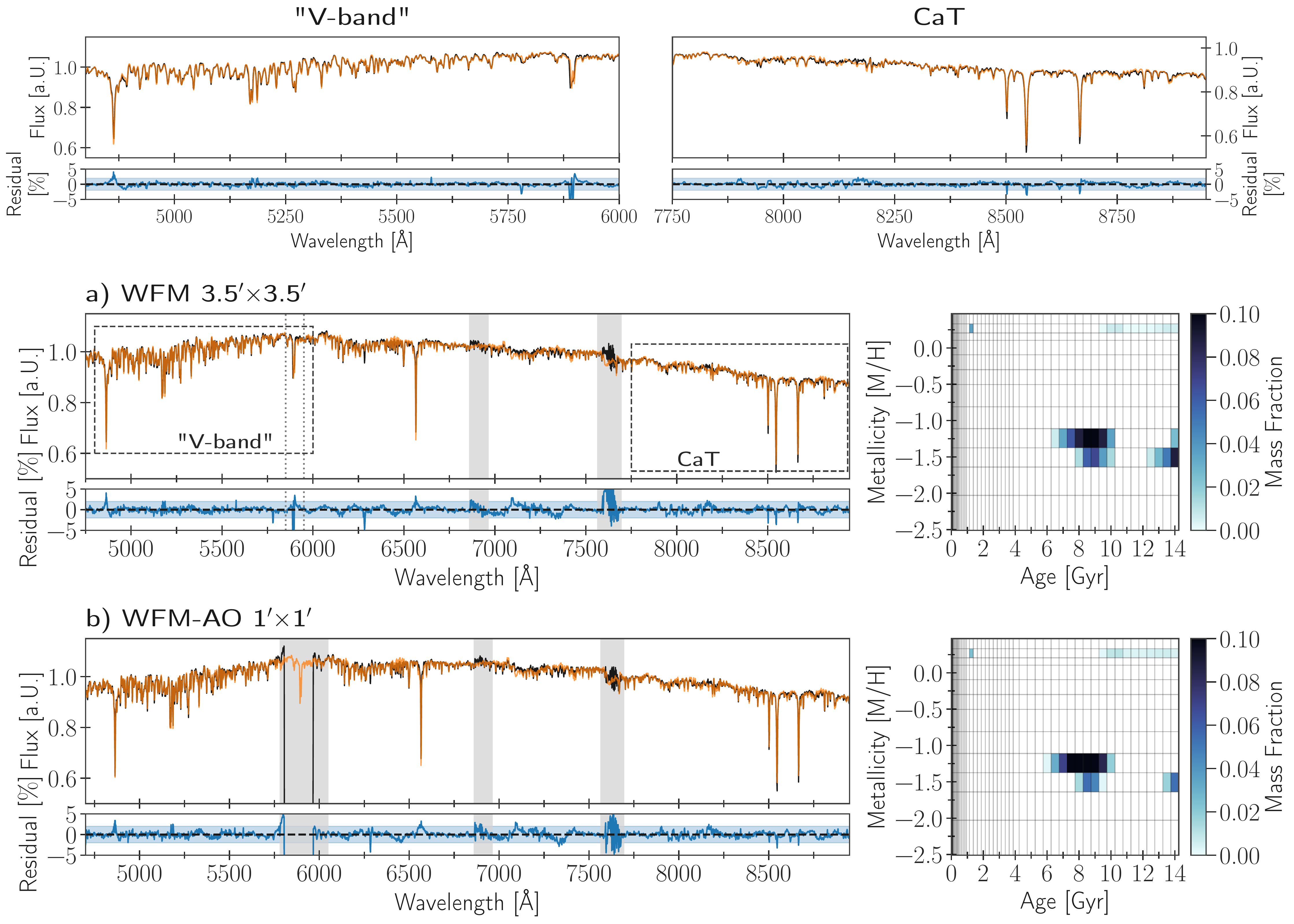}
\caption{\textit{a):} The left panel shows the pPXF fit (orange) to the integrated spectrum of M\,54 (black) from the combined $3.5^{\prime}\times 3.5^{\prime}$ MUSE WFM mosaic. The residuals are shown in blue and the corresponding band shows the 2\% level. The dashed regions are highlighted as zoom-in panels around the ``V-band" and calcium triplet (CaT) region. Grey shaded areas are masked out sky residuals. The grey dotted lines mark the region around the NaD line that was masked out in \protect\citet{mayte} due to interstellar absorption, however the age-metallicity recovery from pPXF is robust against the in- or exclusion of this region. The right panel shows the derived mass fractions in age-metallicity space that make up the bestfit from pPXF. \textit{b):} The same as for the top panel but showing the integrated spectrum from the single pointing MUSE WFM-AO data. We see that the recovery is insensitive to the blocked region from the Na notch filter.}
\label{fig: all_cubes}
\end{figure*}

We then feed these spectra into pPXF with the E-MILES models considering all available ages and metallicities (636 models in total). The results from the pPXF fits are shown in Figure \ref{fig: all_cubes}. We can see that the residuals are on the order of 2\% emphasizing the excellent data and model quality. \par
Both integrated spectra from the $3.5^{\prime}\times 3.5^{\prime}$ MUSE WFM mosaic and the single WFM-AO pointing are almost identical to the eye, and the recovered mass weights in age-metallicity space show a very similar distribution. A quantitative comparison between the recovered mass weights of both data sets is shown in Figure \ref{fig: summary} a) in Appendix \ref{appendix3}.\par
We can identify an old, metal-poor ($\sim1.5$ dex) stellar population at 8 and 14 Gyr, and a young (1 Gyr) and metal-rich (+0.25 dex) contributions. We also see a smaller contribution of old, but metal-rich mass weights. These weights do not fit into our astrophysical picture of chemical enrichment, hence their origin is further explored and discussed in Section \ref{sec: comp}.\par
Here, it is important to see that the lack of the sodium region in the WFM-AO data does not influence the recovery of the stellar populations properties. For the WFM data we also conducted tests of masking and not masking the NaD lines in the same region as \citet{mayte}, as this line is significantly broader than other lines in the spectrum due to interstellar absorption. Both showed consistent results meaning that pPXF is robust against such ``outlier" spectral lines, at least if the wavelength range covers enough other prominent lines. The NaD lines can often be problematic as it is influenced by many different effects, such as the interstellar medium, IMF and sodium abundances. \par
\begin{figure*}[htbp]
\centering
\includegraphics[width=1.5\columnwidth]{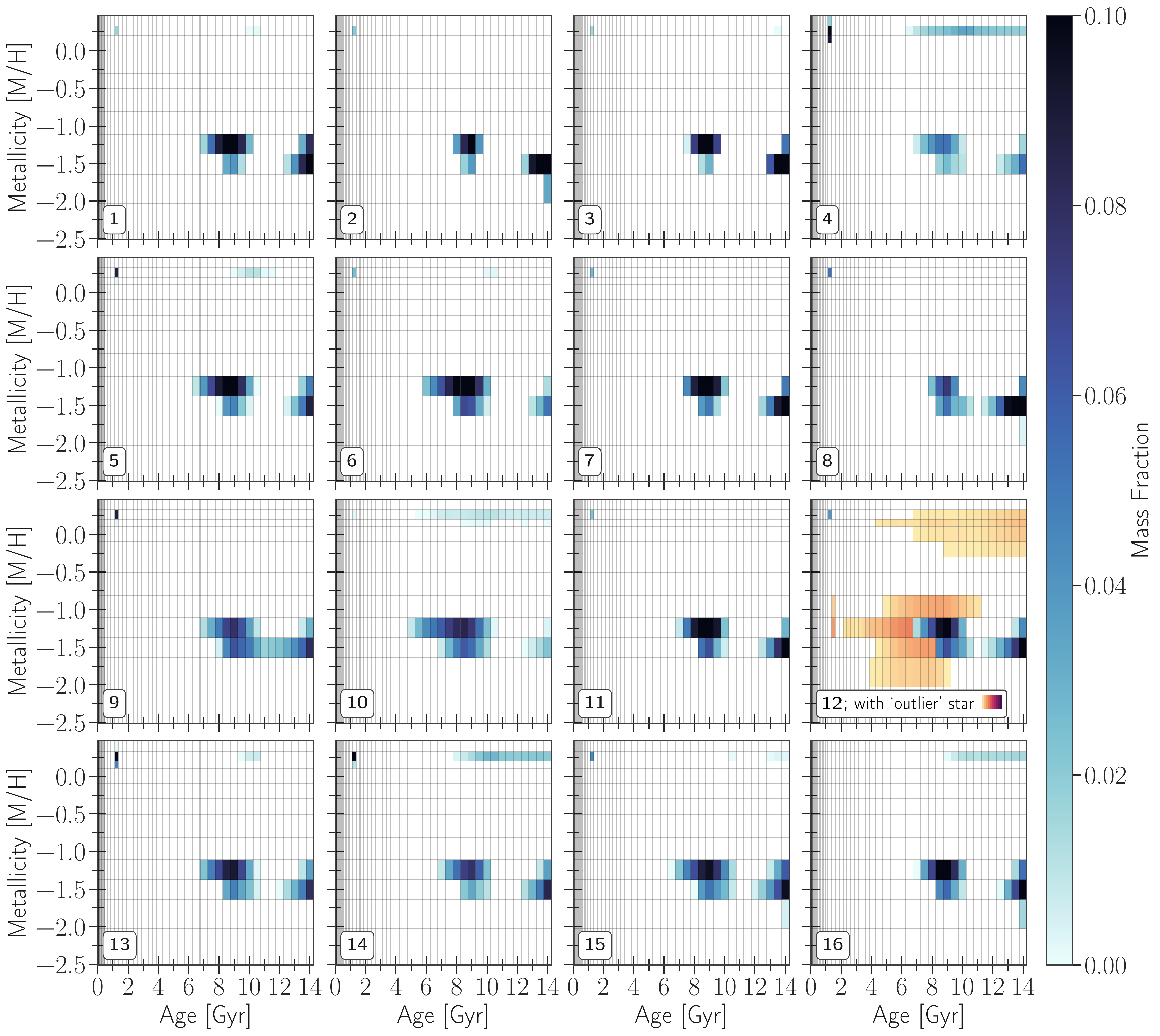}
\caption{Distribution of pPXF-recovered weights in age-metallicity space for the 16 pointings from the MUSE WFM dataset of M\,54 (blue color code). They are arranged in the same order as they appear on the sky in Figure \protect\ref{fig: data}. The red color code (the scale is the same as the blue color code) shows the age-metallicity distribution recovered with the extremely red supergiant star contaminating the spectrum. Note that the apparent presence of an old and metal-rich component is likely related to some of the brightest AGB stars (see Figure \protect\ref{fig: stars_per_cube} and Section \protect\ref{sec: comp}).}
\label{fig: single_cubes}
\end{figure*}

We also show the recovered age-metallicity distribution from the analysis of the integrated spectrum for the 16 single pointings of the MUSE WFM data set in Figure \ref{fig: single_cubes}. Overall, the three populations are picked up in every pointing with varying relative strength (see Figure \ref{fig: summary} b) for a more quantitative comparison). The only significant outlier belongs to pointing number 12 (red color code), where the integrated spectrum is heavily influenced by a red supergiant star ($\mathrm{V}=17.15$ mag \& $\mathrm{I}=13.43$ mag). It can also be clearly identified in Figure \ref{fig: data}, as it is extremely red. This star contributes about 4\% to the total flux of the entire MUSE pointing. If we mask this star out and re-do our integrated light analysis, we obtain a consistent age-metallicity distribution (blue color code) compared to the other pointings.

\subsection{Integrated spectra from individual stars}\label{sec: individ_stars}

The individually extracted stellar spectra from \citet[][see also Section \ref{sec: mayte_analysis}]{mayte} of the same $3.5^{\prime}\times 3.5^{\prime}$ MUSE data set allow us to uniquely combine or exclude certain stars from the integrated spectrum and study the effect on the stellar population recovery. Only stellar spectra with $\mathrm{SNR}\geq 10$ are considered in this sample and they make up roughly 50\% of the total flux from the $3.5^{\prime}\times 3.5^{\prime}$ MUSE mosaic.\par
First, we simply sum up the entire sample of extracted stars (7165 stars) and in a second test case we only consider those stars that were identified as member stars (6656 stars) (see data set C in Table \ref{tab: spectra}). Identical stars that were extracted from different pointings were only accounted for once in the integrated spectrum by taking the corresponding stellar spectrum with the higher SNR. The flux of the individual stellar spectra is preserved and not normalized in any way prior to creating an integrated spectrum from them. \par
\begin{figure*}[htbp]
\centering
\includegraphics[width=\textwidth]{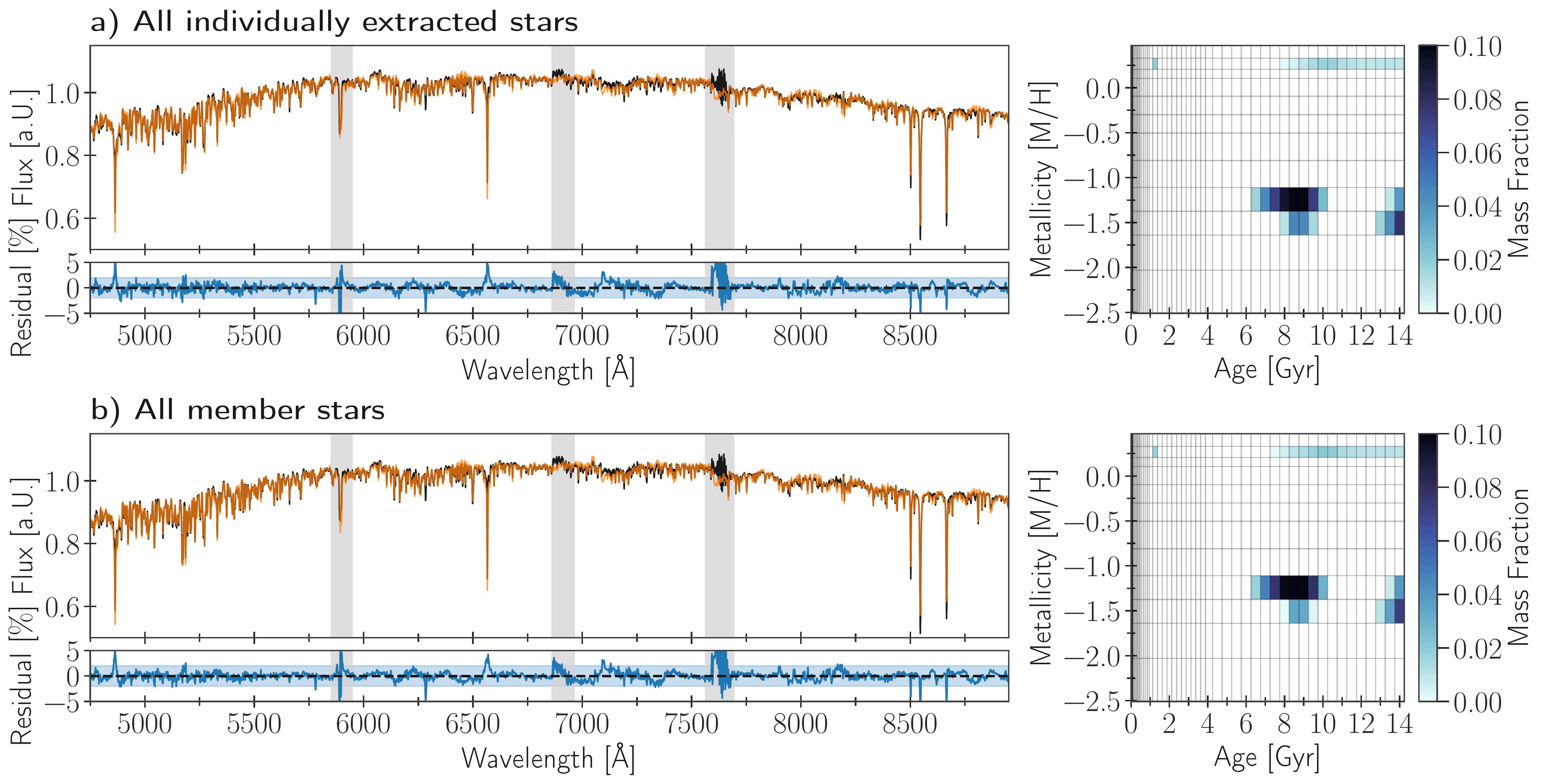}
\caption{\textit{a):} The left panel shows the pPXF fit (orange) to the integrated spectrum of M\,54 (black) from all individually extracted stars \protect\citep[see][]{mayte}. The residuals are shown in blue and the corresponding band shows the 2\% level. Grey shaded areas are masked out regions. The right panel shows the derived mass fractions in age-metallicity space that make up the bestfit from pPXF. \textit{b):} The same as for a) but now showing the integrated spectrum from individual stars identified as \emph{members} of M\,54. Clearly, the stellar population recovery from pPXF is robust against non-member stars.}
\label{fig: single_stars}
\end{figure*}

 
Results from the pPXF fit are shown in Figure \ref{fig: single_stars}. The recovered stellar populations in age-metallicity are nearly identical by eye, which is also shown in a quantitative comparison in Figure \ref{fig: summary} a) in Appendix \ref{appendix3}. Hence, pPXF seems to be robust against contaminating sources, but the stars classified as non-members only make up 8\% of the total integrated light. Still, these non-member stars can have radial velocities of up to -200 km/s, whereas the systemic velocity of M\,54 is around 141 km/s \citep[see Figure 3 of][]{mayte}, which pPXF compensates with a larger velocity dispersion. The fitted velocity dispersion for the integrated spectrum including all extracted stars is around 15 km/s, whereas for the integrated spectrum only including member stars it is 1 km/s, which is the hard coded lower limit in pPXF. The true internal velocity dispersion of M54 is less than the spectral resolution of MUSE. Nevertheless, there is no apparent change in the recovered stellar population properties.

\subsection{Integrated spectra with limiting magnitude cutoffs}\label{sec: cut}

We also look at different magnitude cuts in the CMD and its effect on the recovered stellar populations parameters, as it changes the relative contribution of stars in different evolutionary stages to the total integrated light. This is particularly interesting, if we think that certain regions of the CMD, like for example the horizontal branch, can be responsible for erroneously recovered stellar population properties. Even though making magnitude cuts is inconsistent with a SSP, we expect that the difference becomes negligible at a certain magnitude.  \par
\begin{figure*}[htbp]
\centering
\includegraphics[width=\textwidth]{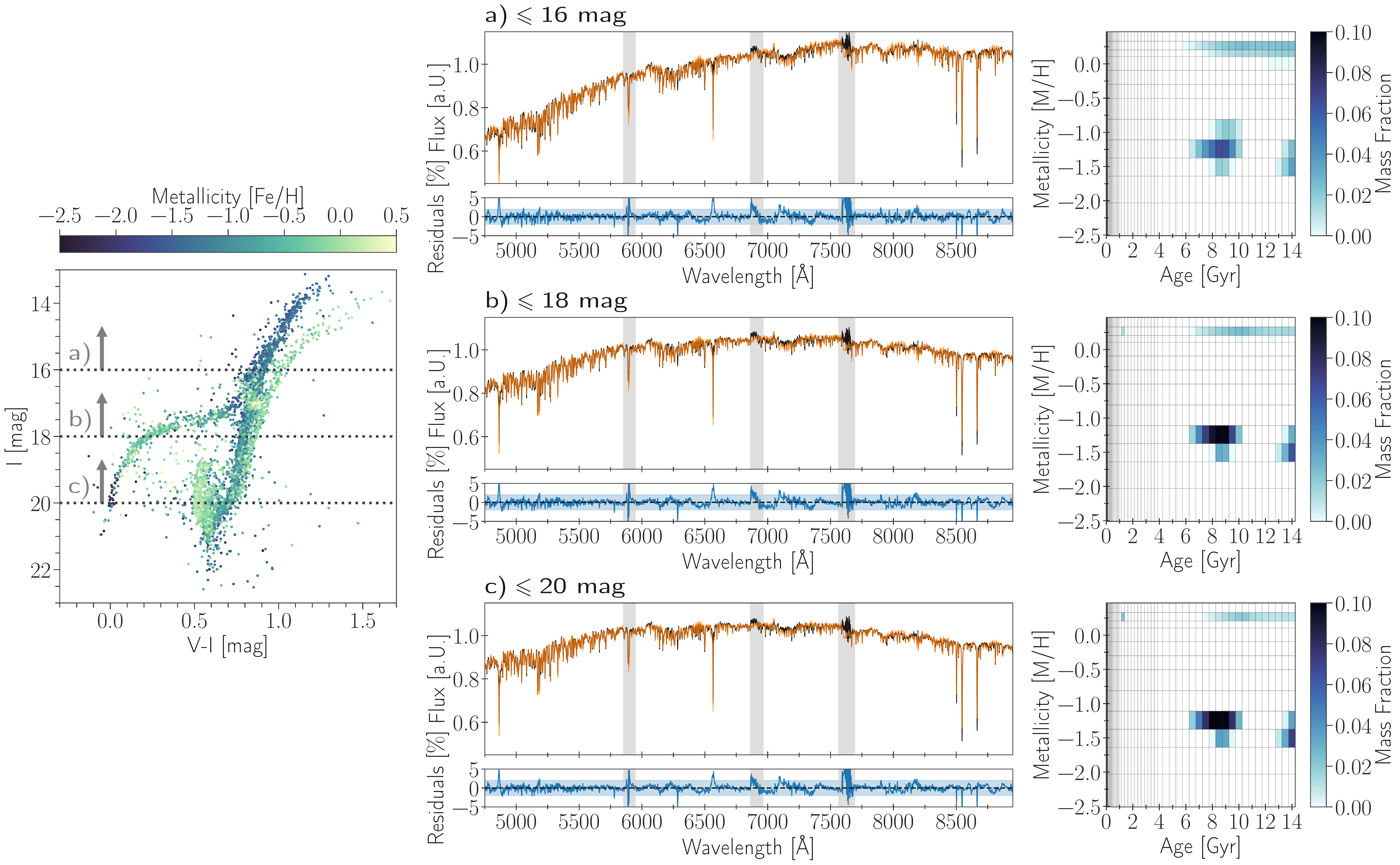}
\caption{\textit{Left}: Color-magnitude diagram of M\,54 member stars color coded by metallicity from \protect\citet{mayte}. The dotted lines show the cuts at 16 (a), 18 (b) and 20 (c) mag from which the three different integrated spectra were made. \textit{Right, top to bottom}: The pPXF fit (orange) to the integrated spectrum of M\,54 (black) from all stars above the magnitude cut is shown for the three cases a), b) and c). The residuals are shown in blue and the corresponding band shows the 2\% level. Grey shaded areas are masked out regions. The corresponding derived mass fractions in age-metallicity space that make up the bestfit from pPXF are also shown. Already for $\mathrm{I}\leq 18$ mag, the same distribution is found as if all member stars were included in the integrated spectrum.}
\label{fig: magnitude_cut}
\end{figure*}

We show in Figure \ref{fig: magnitude_cut} three different cases, where we include all member stars with I-band magnitudes brighter than 16, 18 and 20 mag respectively (see data set E in Table \ref{tab: spectra}). These cuts encompass 53\%, 89\% and 98\% of the total integrated light of all member stars. \par 
For the first case, a), only stars brighter than the red clump are considered, for which pPXF still finds the old and intermediate-age, metal-poor population, but no longer the young, metal-rich one. This is likely due to the fact, that the fraction of young, metal-rich stars contributing to the total spectrum is lower in this particular space of the CMD. As a consequence, there are more weights in the old ($> 8$ Gyr), metal-rich ($> 0.0$ dex) age-metallicity regime. \par
For the other two cases, b) \& c), the result becomes identical to Figure \ref{fig: single_stars}, where all member stars were taken into account. This can also be seen in Figure \ref{fig: summary} c) in Appendix \ref{appendix3}. In particular for case b), pPXF is able to reproduce the same results, \emph{as if} the data had a limiting magnitude of 18 in the I-band, which is just below the red clump. This is not surprising, as the magnitude cut at 18 already includes 89\% of the total integrated light coming from all member stars. \par
Moreover, it is reassuring to see that relatively larger contribution from the horizontal branch to the total integrated spectrum in b) does not artificially induce any additional young populations. This is investigated further in the following section.

\subsection{A closer look at M\,54's horizontal branch}\label{sec: hb}

M\,54 has quite an extended horizontal branch (HB) with a ratio of 0.75 \citep[][a HB ratio of 1 and -1 means only blue or red HB stars respectively]{georgiev09}, which has been argued to bias age determinations by about 2 Gyr or more \citep[e.g.][]{lee00,schiavon04,colucci09,georgiev12} or to cause spurious young populations in integrated light analysis \citep[e.g.][]{ocvirk10}. In particular, metal-poor globular clusters have been found to exhibit bluer horizontal branches, which have strong Balmer lines and hence can mimic young main sequence stars. Generally, HB stars are difficult to model in SSP models, as there are higher order parameters determining their morphology \citep[e.g.][]{lee94,gratton10}. \par
In principle, the extended HB in M\,54 could be the reason that we find the metal-poor population ($\sim$ -1.5 dex) at 8 Gyr even though we know that it is likely older than that. The advantage of our MUSE data set is that we can test this hypothesis by constructing an integrated spectrum \emph{without} M\,54's HB stars and fit it with pPXF. We exclude all stars that either have $\mathrm{I}<18.5$ mag and $\mathrm{V-I} <0.7$ mag or $\mathrm{I}>18.5$ mag and $\mathrm{V-I} <0.3$ mag. \par
\begin{figure*}[htbp]
\centering
\includegraphics[width=\textwidth]{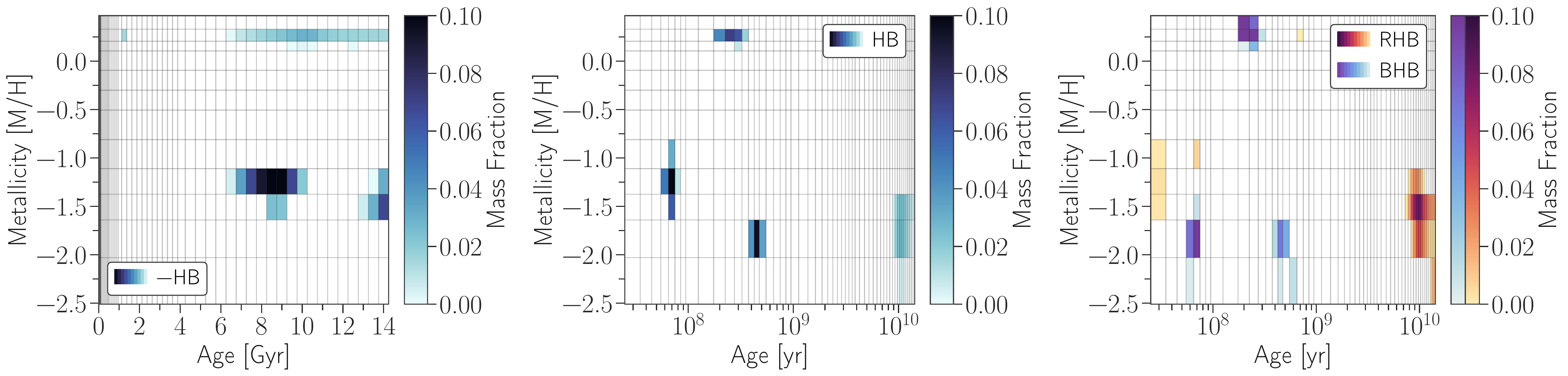}
\caption{\textit{From left to right:} pPXF recovered mass weights in age-metallicity space from fitting E-MILES models to the integrated spectrum of M\,54 made from all member stars \emph{without} the horizontal branch (HB), only HB stars, only red and blue HB stars respectively.}
\label{fig: horizontal_branch}
\end{figure*}

As can be seen from Figure \ref{fig: horizontal_branch} the recovered stellar populations properties are identical to the fit, where all members stars were included in the integrated spectrum (see Figure \ref{fig: summary} c). Hence, we can conclude that the HB is not responsible for potentially shifting the metal-poor population to 8 Gyr. In Section \ref{sec: results} we give another possible explanation for this being due to the large oxygen abundances in M\,54. \par
On the other hand, if we fit the integrated spectrum made of \emph{only} HB stars, we indeed recover very young ($< 1$ Gyr), metal-poor and metal-rich alike, populations and an old ($\sim 10$ Gyr), metal-poor population. We can even attribute the recovered old component to red ($\mathrm{V-I} >0.43$ mag) and the very young components to blue ($\mathrm{V-I} >0.43$ mag) HB stars. \par
Interestingly, pPXF always recovers a very small mass fraction ($<1 \%$, which corresponds to $<10 \%$ in light) in the youngest age bin (0.03 Gyr) from fitting the E-MILES models to the MUSE integrated spectrum with \emph{and} without HB stars alike. Hence, we cannot attribute this spurious weight in the youngest age bin to the presence of blue HB stars in the integrated spectrum as was found in \citet{ocvirk10}\footnote{When SSP models with ages younger than 1 Gyr are excluded from the fit, this systematic vanishes, therefore it can likely be attributed to uncertainties in the SSP models at these young ages.}.

\subsection{Bluer wavelengths from WAGGS}\label{sec: waggs_analysis}

The influence of hot stars, such a young stars or evolved horizontal branch stars, starts to become dominant in bluer wavelengths ($< 4000$ \AA) than the MUSE range. To be conclusive that M\,54's extended HB does \emph{not} influence the age-metallicity distribution recovery as found in the previous section, we here analyze M\,54's integrated spectrum from the WAGGS survey until 3500 \AA\footnote{Below that wavelength the spectrum becomes noise dominated.} (see data set F in Table \ref{tab: spectra}). \par
\begin{figure*}[htbp]
\centering
\includegraphics[width=\textwidth]{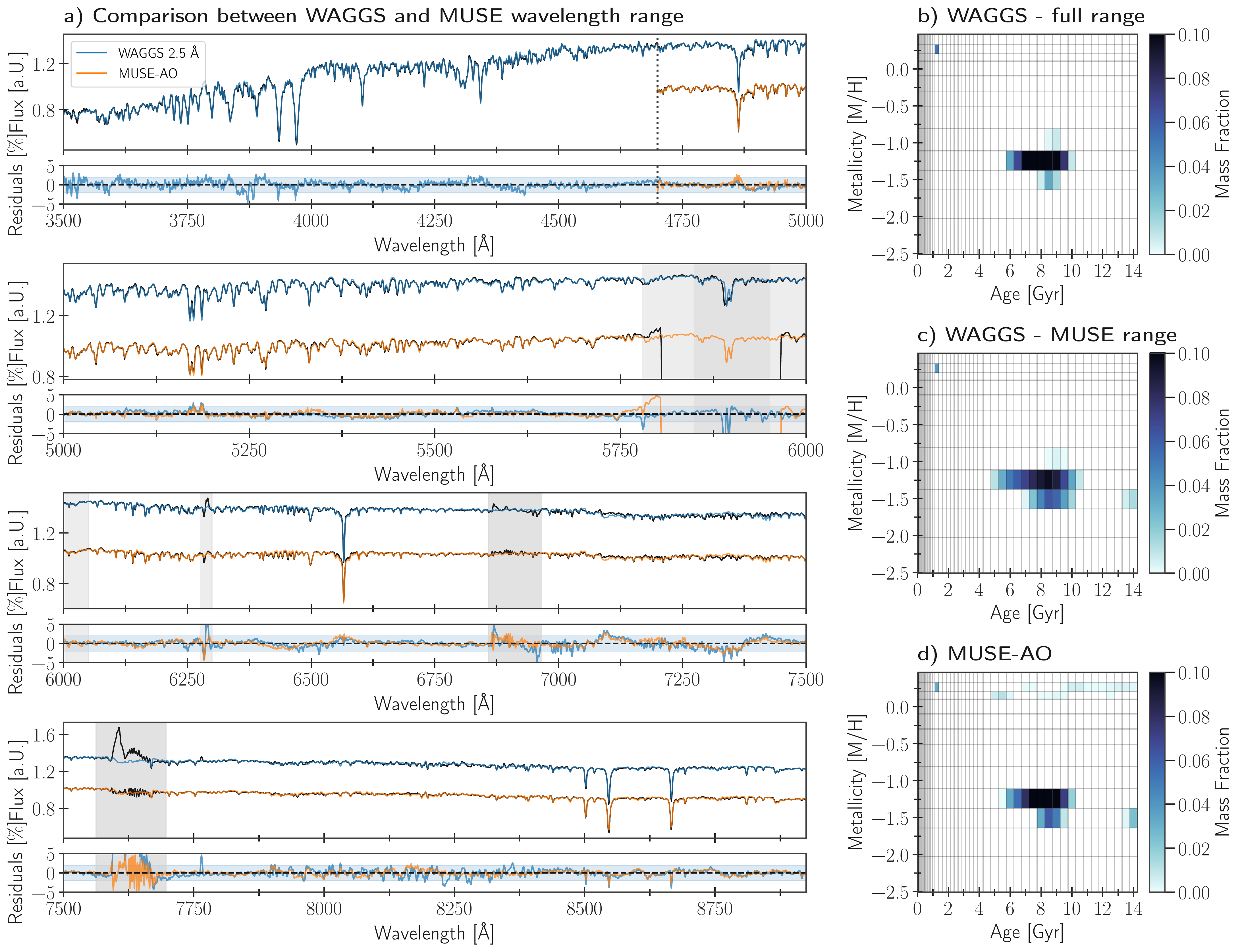}
\caption{\textit{a):} Comparison of pPXF fit to the integrated spectrum of M\,54 between WAGGS (blue) and MUSE-AO (orange). The WAGGS spectrum was broadened to a FWHM of 2.5 \AA, and the MUSE spectrum was made from the same field-of-view as WiFes. The blue band shows again the 2 \% level of the residuals. Both residual spectra are very similar emphasizing the high data quality. \textit{b):} Recovered age-metallicity distribution from the full wavelength range of the broadened WAGGS spectrum. \textit{c):} Same as b), but now only considering the MUSE wavelength range (see dotted line in top panel of a). \textit{d):} Recovered age-metallicity distribution from the MUSE-AO spectrum with same field-of-view as WiFes.}
\label{fig: waggs1}
\end{figure*}

In Figure \ref{fig: waggs1} a) we show the pPXF fits to the entire WAGGS spectrum and the integrated spectrum from the MUSE-AO observations. We chose the latter for the comparison as it is centered on M\,54 and hence could easily be cropped to the same field-of-view as WiFes in order to eliminate changes in the stellar population recovery induced by the differences in the spatial coverage. It is reassuring to see that both spectra acquired with completely different instruments are almost indistinguishable. The same applies to the residuals of pPXF, also when the WAGGS spectrum is fitted at native resolution (see Figure \ref{fig: waggs2} in Appendix \ref{appendix2}).\par
The recovered mass weights in age-metallicity space fitted to the whole wavelength range of WAGGS, only the wavelength covered by MUSE and to the actual MUSE-AO observations are plotted in Figure \ref{fig: waggs1} b), c) and d) respectively (see also Figure \ref{fig: summary} d) in Appendix \ref{appendix3} for a direct comparison). Including these bluer wavelengths does not recover any artificial young populations below 1 Gyr, therefore we conclude that the horizontal branch stars have no influence on the stellar population recovery from the integrated light in our analysis of M\,54. Quite contrary, pPXF now puts all the mass weights at the 1 and 8 Gyr old population and the one at 14 Gyr vanishes. It reappears though, if the MUSE wavelength range is considered with the WAGGS spectrum. This could either imply that we loose the ability to recover very old ages, if we include these blue wavelengths or that the 14 Gyr population is not real/robust. However, it is more likely that this is associated with the overall difficulty to distinguish between SSP models with ages of 8 Gyr and above at fixed metallicity. Therefore, the ``two" populations at 8 and 14 Gyr could also be just one. In Figure \ref{fig: waggs2} in Appendix \ref{appendix2} we show M\,54's integrated spectrum from WAGGS and MUSE fitted by the PEGASE-HR models, suggesting a more extended old, metal-poor population.

\subsection{The influence of the brightest stars}\label{sec: comp}

\begin{figure*}[htbp]
\centering
\includegraphics[width=1.5\columnwidth]{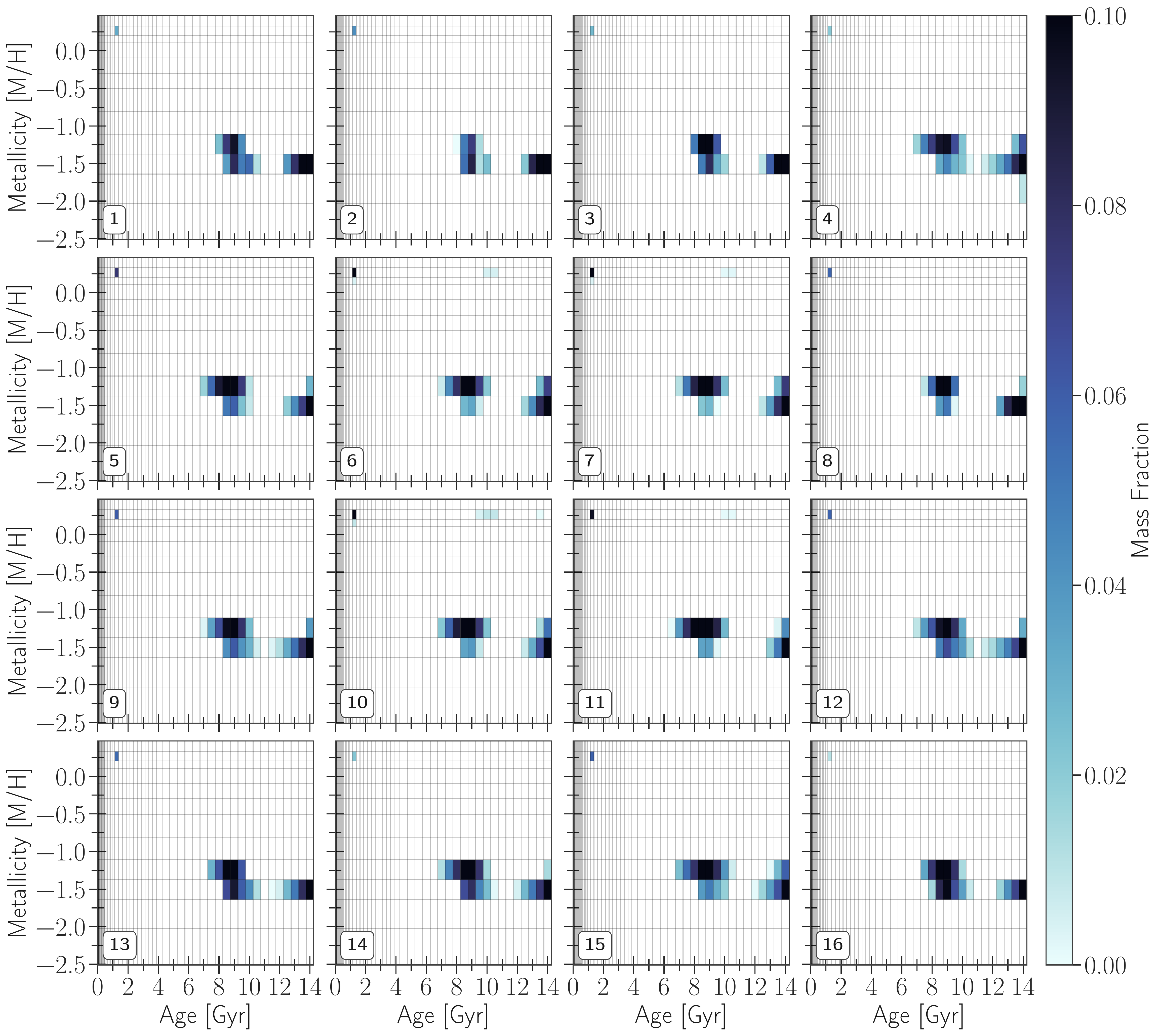}
\caption{Distribution of pPXF-recovered weights in age-metallicity space for member stars of M\,54 associated with the MUSE pointings they were extracted from. They are arranged in the same order as they appear on the sky in Figure \protect\ref{fig: data}. In pointings number 6, 7, 10, 11, 12 and 16 bright ($\mathrm{I}<14$ mag) and extremely red stars ($\mathrm{V-I}>1.7$ mag) are not included in the final integrated spectrum (see Section \protect\ref{sec: individ_stars} \& Appendix \protect\ref{appendix4}).}
\label{fig: stars_per_cube}
\end{figure*}

The recovery of an old ($8-14$ Gyr) and metal-rich (+0.25 dex) component in all of the above age-metallicity distributions does not necessarily fit into our astrophysical understanding of chemical enrichment in the presence of the three other populations. We therefore investigated, whether this component is real. From Figure \ref{fig: single_stars} it seems that this component might arise from a few individual pointings. To explore this further, we repeated the exercise by only considering the classified member stars of M\,54. We associate each member star of M\,54 to the MUSE pointing it was extracted from, create an integrated spectrum and fit it with pPXF. For pointing number 12 we again excluded the red supergiant star as before.\par
Overall, the three populations, the old, metal-poor ($\sim1.5$ dex) at 8 and 14 Gyr, and the young (1 Gyr), metal-rich (+0.25 dex), are much more clearly recovered as seen in Figure \ref{fig: stars_per_cube} (see also Figure \ref{fig: summary} e) for a more quantitative comparison). The old ($8-14$ Gyr), metal-rich (+0.25 dex) component in pointing numbers 4, 5, 13 and 14 from Figure \ref{fig: single_cubes} vanishes. Looking at those fits again, it is evident that foreground stars with large offsets in their line-of-sight velocities are disturbing the integrated spectrum and the fit when summing up the full MUSE cubes. \par
However, this is not the case for the central pointings (6, 7, 10 and 11) and number 16. In fact, only including M\,54's member stars in the integrated spectra makes the old, metal-rich population much more prominent in the central four pointings. We speculate that this could be due to two effects, a population of high-metallicity MW foreground stars or perhaps thermally pulsing AGB stars. The former possibility was tested by running the Besan\c{c}on Milky Way model \citep{besancon2}\footnote{\url{https://model.obs-besancon.fr/modele_home.php}} predicting around $100-150$ MW stars still left in the M\,54 member star sample. These stars have preferably old ages and solar metallicities, however the ages and metallicities of the individual M\,54 member stars from \citet{mayte} did not yield any measurements in that parameter range (see also Section \ref{sec: results}). \par
To explore the second possibility, we looked at the color-magnitude diagram of the member stars per MUSE pointing (see Figure \ref{fig: cmd_per_cube} in Appendix \ref{appendix4}). Especially in the four central pointings, brighter ($\mathrm{I}<14$ mag) and much redder ($\mathrm{V-I}\gg1.7$ mag) stars are found, that could be thermally pulsing asymptotic giant branch (AGB) stars. These brightest stars contribute around 20\% of flux to the total integrated spectrum and therefore have a non-negligible influence on the shape and spectral features of the integrated spectrum. Their red continuum shape and typical spectral features like prominent TiO bandheads can easily be mimicked by old, metal-rich stellar populations, if they are not properly accounted for in the SSP models \citep[see e.g.][for the influence of AGB stars]{maraston05,maraston06}. If they are indeed the source of an old, metal-rich component then it might also explain why the component was \emph{not} recovered from the WAGGS data (see Section \ref{sec: waggs_analysis}), as we expect that the influence of AGB stars becomes weaker at bluer wavelengths.\par 
Excluding these stars from the integrated spectrum of the four central pointings and repeating our analysis, we obtain the age-metallicity distributions as shown in Figure \ref{fig: stars_per_cube}. The contribution of the old, metal-rich component decreases significantly from $20-30\%$ to $0.3-2\%$. Evidently, the old, metal-rich component is not recovered in a significant amount when considering the integrated spectrum made from the full MUSE cube in three out of the four central pointings (Figure \ref{fig: single_cubes}). We attribute this to the fact that the full cubes contain enough flux from fainter, unresolved stars, such that the contribution of the brightest stars drop to around 10\%.\par
For pointing number 16 a single star (marked by a grey circle in Figure \ref{fig: cmd_per_cube}) is responsible for the recovery of the old, metal-rich component. This star also appears in pointing number 15, but its presence in the integrated spectrum of this pointing does not cause the old, metal-rich component to appear - probably because its flux contribution is only 4\%, whereas in pointing number 16 it is 8\%.\par
Counterintuitivly, the recovered age-metallicity distributions for pointings 4, 5 and 8 in Figure \ref{fig: stars_per_cube} do not show the old, metal-rich component even though their integrated spectra \emph{include} stars with $\mathrm{I} > 14$ mag that contribute around 20\% to the total flux. Nevertheless, we confirm that excluding these brightest stars from the integrated spectra from Figures \ref{fig: single_stars} b) and \ref{fig: magnitude_cut} b) \& c), makes the old, metal-rich component completely vanish, while for Figures \ref{fig: magnitude_cut} a) and \ref{fig: horizontal_branch} the contribution significantly decreases. Instead, the relative contribution of the young, metal-rich component becomes stronger in all cases. \par

\section{Integrated vs. resolved age-metallicity distribution recovery}\label{sec: results}

\begin{figure*}[htbp]
\centering
\includegraphics[width=\textwidth]{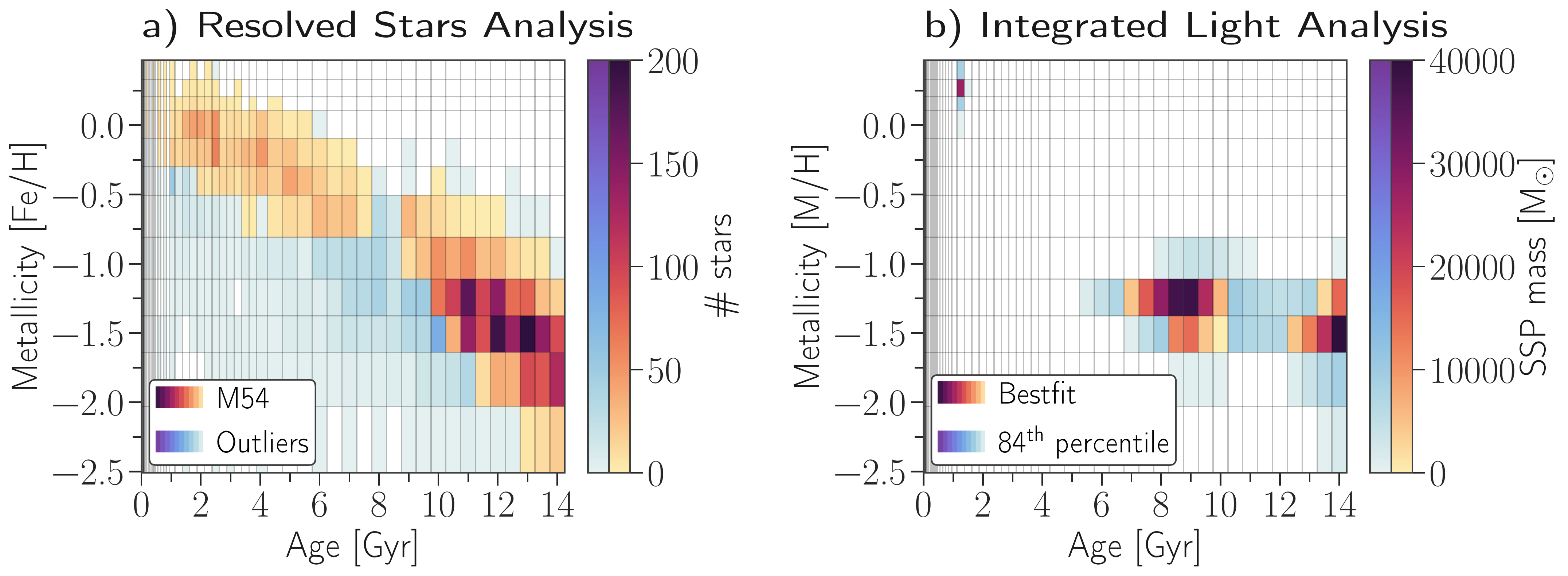}
\caption{\textit{a):} Number of resolved stars in their respective age-metallicity bin as determined in \protect\citet{mayte}. The bins correspond to the age-metallicity grid adopted by \protect\citet{vazdekis16} using the BaSTI isochrones \protect\citep{basti04}. The red color code shows the stars that were characterized to M\,54, whereas the blue color code shows outliers. The mean age and metallicity are 9.24 Gyr and -1.01 dex respectively. \textit{b):} Age-metallicity distribution from fitting the full, integrated spectrum of M\,54. Here, we show the result for the integrated light of individual member stars spectra \emph{excluding} the brightest stars (see Section \ref{sec: comp}). The colorbar indicates now how much absolute stellar mass is contained in each SSP bin. The red color code shows the distribution of mass bins belonging to the bestfit solution, whereas the blue color code shows the extent of the 84th percentile mass bins derived from randomly re-sampling the residuals. The mass-weighted mean age and metallicity are 9.53 Gyr and -1.21 dex respectively.}
\label{fig: comparison}
\end{figure*}

As we have seen in the previous section, the recovery of the mass distribution in age-metallicity space does not heavily depend on the exact approach used to construct the integrated spectrum of M\,54 - however the contribution of the brightest stars is possibly responsible for the recovery of unphysical mass weights for this particular system (Section \ref{sec: comp}). Therefore, we restrict our comparison to the resolved stellar population analysis from \citet{mayte} to the results from the integrated spectrum built from M\,54's members \emph{excluding} the brightest stars (see Section \ref{sec: comp}). \par
In Figure \ref{fig: comparison} a) we show the age-metallicity relation derived from the single stars binned to the same age-metallicity grid as the one we use in our integrated analysis to ease comparison. We show the stars that belong to M\,54 in a red color code, where as stars that have been characterized as outliers in blue color-code, as quantified by Gaussian mixture models in \citet{mayte}. It is also important to note again that the horizontal branch stars are not included in this result. \par
In Figure \ref{fig: comparison} b) we show our result from the integrated light analysis. Now, the color code shows the absolute mass contained in each SSP bin instead of the relative mass fraction (see equation \ref{eq: mass}). We show the bestfit mass bins as well as the 84th percentile from randomly re-sampling the residuals from pPXF, adding them to the bestfit spectrum and re-fitting 100 times. We chose to keep the regularization parameter fixed to the value derived for the bestfit, as opposed to use no regularization at all. This allows for studying the effect on the smoothening of individual mass bins in age-metallicity space purely due to random variation in the fitted spectrum, and not the change of the regularization parameter. We also use the variation of the derived mass fraction from the 16 nearly independent MUSE pointings to quantify how much their absolute value as opposed to their smoothening across the age-metallicity plane. The mean relative differences between the mass fractions recovered for each pointing and the total of all member stars is around 28\% and 38\% for age and metallicity respectively (see Figure \ref{fig: summary} e)). They are on the same order as the ages and metallicities measured from the individual stars from \citet{mayte}, which vary by 44\% and 26\% respectively across the pointings. They are likely induced by various factors such us differing SNR of the integrated spectrum or stochastic sampling of certain stars. \par
By focusing on the red color code of Figure \ref{fig: comparison}, we see that the resolved star analysis shows a nicely rising chemical enrichment as a function of time. Most stars are roughly between 10 and 14 Gyr in age and -2.0 to -1.0 dex in metallicity. A more spread-out population of stars lie between 3 and 5 Gyr and around -0.25 dex in metallicity, whereas more stars seem to concentrate again at around 2 Gyr and solar metallicity. \par
The integrated light analysis on the other hand shows three more concentrated populations at 14 Gyr and -1.5 dex, around 8 Gyr and -1.25 dex and at 1 Gyr and super solar metallicity respectively. The returned chemical enrichment is also less continuously rising, but shows a rather flat enrichment from 14 to 8 Gyr and then jumps to +0.25 dex in metallicity at 1 Gyr. However, the separation between the 8 and 14 Gyr is likely not real, but either induced by the general poor age resolution at old ages, or by the quite complex individual element abundances of the Sagittarius nucleus. As shown by \citet{carretta10a,carretta10b}, stars in M\,54 exhibit almost a one dex spread in oxygen abundance. This effect can bias age determinations between $5-10\%$ towards younger ages for generally old and iron-poor stars \citep{vandenberg12}, as the turn-off region becomes bluer for enhanced oxygen. In fact, the same phenomenon was observed in the resolved study \citep[][who also recover a population around 8 Gyr from stars of the Sagittarius stream]{deboer15}, however by applying a Gaussian age prior the authors could eliminate this problem \citep[see][Section 3.4]{mayte}. \par
Similarly, non-solar $\alpha$-abundances (especially magnesium) have a strong effect on the position of the red giant branch. It is well known that the Sagittarius stream as well as its nucleus, M\,54, follow the well defined [$\alpha$/Fe]-[Fe/H]-relation \citep[see][]{carretta10a,deboer14,mucciarelli17}. Stars with an iron abundance around -1.5 dex are $\alpha$-enhanced by about 0.4 dex, whereas stars with [Fe/H]-ratios between -1.0 and +0.0 dex have [$\alpha$/Fe]-ratios between +0.2 and -0.2 dex respectively. The E-MILES models used in our integrated analysis are based on the ``baseFe" SSP models, which refers to the assumption that the MILES stars have solar $\alpha$-abundances (i.e. $\mathrm{[Fe/H]}=\mathrm{[M/H]}$). However, these stars actually follow the MW [$\alpha$/Fe]-[Fe/H]-relation \citep[as determined in][]{milone11} and hence the solar-scaled BaSTI isochrones are inconsistent with low metallicity MILES stars. Unfortunately, this likely does not explain why the metallicity derived for the metal-poor population in the integrated analysis is consistent with the resolved study and the young population is super-solar compared to solar values from the individual M\,54 stars. If $\alpha$-abundances are causing the metallicities differences between the two methods, we would expect to observe the opposite trend. Furthermore, the fit residuals of the Mgb lines (see e.g. Figure \ref{fig: waggs1}) show that the magnesium abundance is actually overpredicted in the SSP models suggesting discrepancies between the adopted and actual alpha-abundance of M\,54. \par
Furthermore, the discontinuity between the old, metal-poor and the young, metal-rich population in the integrated analysis can have a number of possible explanations. It could be that the relative contribution of stars between 2 and 8 Gyr as seen in the resolved analysis is not significant in the integrated spectrum to be picked up by pPXF. This means that the star formation rate of a certain star forming episode has to reach a specific threshold to be contributing significantly to the integrated light. It could also be that this is an issue of how regularization is applied, as it smooths the mass weights only in the horizontal and vertical direction in the age-metallicity plane, but not diagonally. \par
Despite the apparent differences (or similarities) between the age-metallicity distributions of the resolved and integrated light analysis, the average quantities of both methods agree well. We quote a mean age and metallicity of 9.24 Gyr and -1.01 dex for the resolved stars and a \emph{mass-weighted} mean age and metallicity of 9.53 Gyr and -1.21 dex for the integrated method. This corresponds to a difference of only 3\% in age and 0.2 dex in metallicity, which is good precision considering the range of metallicities of almost 2 orders of magnitude. Weighting the resolved stars by their V-band luminosity yields a mean age of 9.69 Gyr and metallicity of -1.11 dex, whereas light-weighted quantities from the integrated method produce a mean age of 7.34 Gyr and metallicity of -0.99 dex.\par
The averages across the 16 MUSE pointings vary by about 0.43 Gyr and 0.06 dex for the resolved and by 0.82 Gyr and 0.10 dex for the integrated analysis. Statistical errors on these quantities from both \citet{mayte} and our random re-sampling of the residuals are below a few percent. However, based on the general uncertainty of stellar population synthesis as well as the poor age resolution at old ($\gtrsim 8$ Gyr) ages, we do not claim to recover the \emph{true} mean age and metallicity of M\,54 to better than 20\%. \par
To summarize, the integrated light analysis of M\,54 can recover a young, metal-rich and old, metal-poor stellar population even though pPXF is free to choose any, not necessarily physical, age-metallicity combination that best represents the observed integrated spectrum. The derived mean ages and metallicities are consistent with the resolved analysis despite the differences in the used SSP models and the lack of considering individual element abundances that are present in M\,54.

\subsection{Mass-to-light ratios from integrated analysis}

\begin{figure}[htbp]
\centering
\includegraphics[width=\columnwidth]{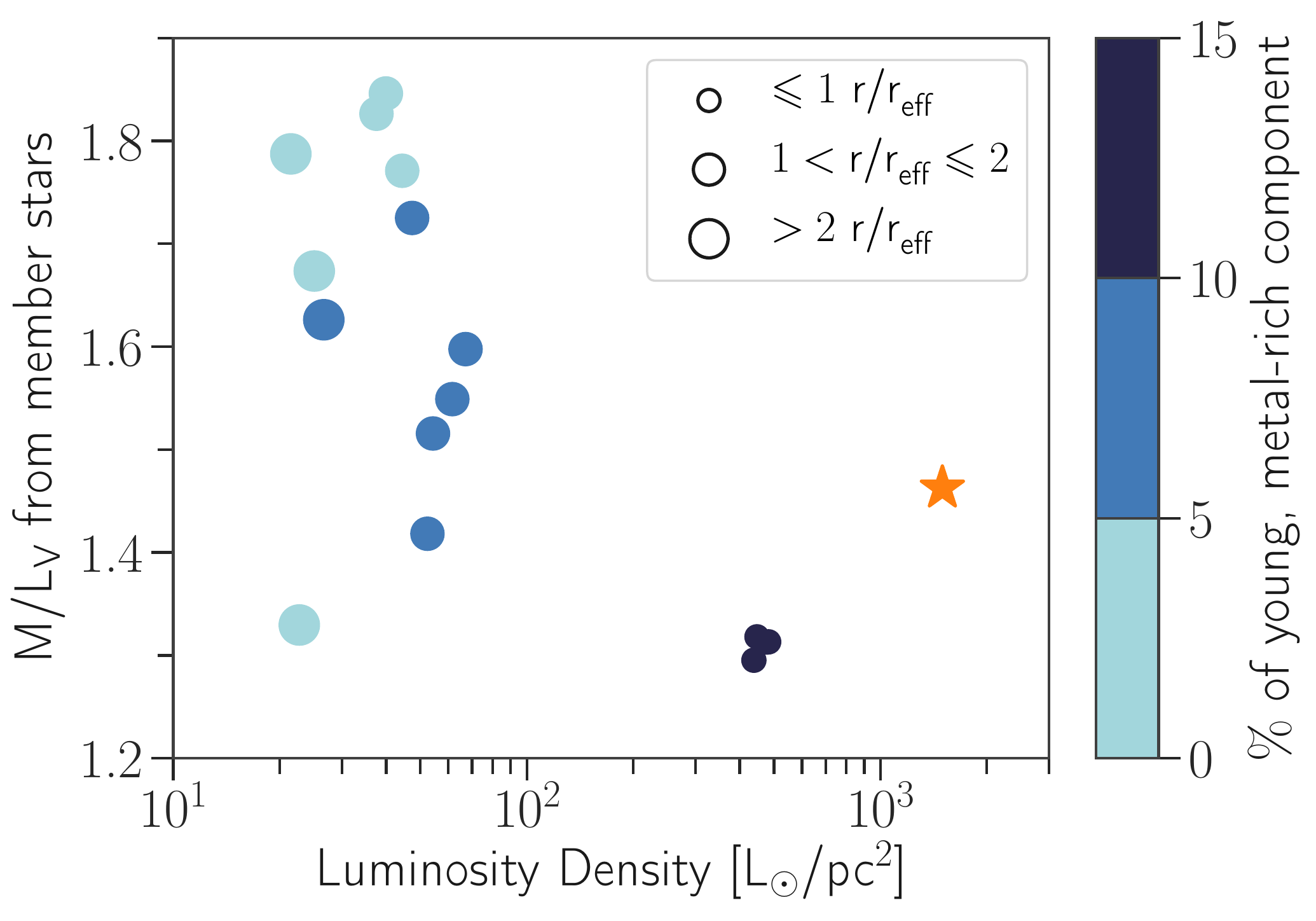}
\caption{Mass-to-light ratios in the V-band derived from returned pPXF mass weights for integrated spectra made from M\,54's member stars excluding the brightest stars (see Section \protect\ref{sec: comp}) plotted against the luminosity density in each corresponding pointing. The symbol size corresponds to the distance from M\,54's center, while the color-code shows the fraction of the young (1 Gyr), metal-rich (+0.25 dex) component. The orange star corresponds to the global M/L ratio obtained from an integrated spectrum made from \emph{all} the M\,54's member stars excluding the brightest stars. Estimated errors from randomly re-sampling the residuals is on the order of the symbol size.}
\label{fig: mlratio}
\end{figure}

From the returned pPXF mass weights we also calculate stellar M/L ratios in the V-band (see equation \ref{eq: ml}) for integrated spectra made from M\,54 member stars (excluding the brightest stars; see Section \ref{sec: comp}). This is done for our canonical E-MILES SSP library choice with a bimodal IMF of slope 1.3. We find a global value M/L$_{\mathrm{V}}=1.46$, however across the 16 MUSE fields we find values that vary from 1.3 to 1.8 (or with a standard deviation of 0.22 from the global M/L). These are shown in Figure \ref{fig: mlratio} as a function of luminosity density color-coded by contribution of the young (1 Gyr), metal-rich (+0.25 dex) component. From this it appears that the four central pointings, which have the highest luminosity density, tend to have lower M/L ratios. The outer fields at lower luminosity density show a wide spread in M/L ratios. We confirm that the scatter in the derived M/L ratios originate from the varying relative contribution of the young, metal-rich mass fractions, as this directly translated to a change in the M/L ratio (see equation \ref{eq: ml}). This agrees with the young, metal-rich stars being more centrally concentrated as found by \citet{mayte}. There is an outlier corresponding to pointing number 4, which has a low M/L ($\approx 1.3$) at low luminosity density ($\approx20\ \mathrm{L}_{\odot}/\mathrm{pc}^2$), likely caused by the brightest star contributing around 25\% to the total flux (see Figure \ref{fig: cmd_per_cube}). \par
The exclusion of the horizontal branch in the composite spectrum (see Section \ref{sec: hb} after \ref{sec: comp}) yields a M/L ratio of 1.45, whereas the magnitude cuts (see Section \ref{sec: cut} after \ref{sec: comp}) yield 1.53, 1.38 and 1.48 for $\mathrm{I}<16$, 18 and 20 mag respectively. Even though these changes are below our estimated statistical errors of $1-2\%$, we do not claim that they are significant, especially with respect to typical uncertainties of 6\% in other studies \citep[e.g.][]{cappellari13}. \par
Our stellar population derived M/L$_{\mathrm{V}}$ ratio agrees with measurements from \citet{kimmig15}, who modelled M\,54's velocity dispersion profile with a King profile and taking internal rotation into account. However, it is lower by about 0.5 compared to studies from \citet{baumgardt18,waggs3}, who fitted N-body simulations without internal rotation to the velocity dispersion profiles. This might be an indication that it is important to include internal rotation in the dynamical modelling, as it decreases the M/L ratio measurement. There is evidence after all that M\,54's young, metal-rich population exhibits a significant rotation signature \citep{bellazzini08,mayte2}.

\section{Discussing comparisons of integrated and resolved studies}\label{sec: discussion}

After thoroughly having compared the stellar population results from resolved star analysis of \citet{mayte} and our integrated light analysis in the previous section, we have now arrived at the question, whether this comparison proves that the recovery of age-metallicity distributions from integrated spectra provides us with the same information content as the resolved stars with regard to stellar population ages and metallicities. \par
Assuming now that the derived ages and metallicities of the resolved stars resemble the `best' knowledge we have about M\,54, one might take the mismatch between the two approaches in Figure \ref{fig: comparison} as a failure of the integrated method. However, the importance and success of this comparison is not to be measured in how perfectly the individual bins in age and metallicity match each other, but the fact that the integrated light analysis can clearly and robustly detect that M\,54 hosts \emph{multiple} populations, which are \emph{even} located in a similar age-metallicity space as the properties of the resolved stars. Considering that the resolved and integrated analysis techniques are also very different conceptually and in the models they use, the similarities of the recovered age-metallicity distributions are compelling. \par
In the following sections we will discuss in more detail, which aspects of stellar population analysis have to be considered to perform a one-to-one comparison between the resolved and integrated methods (Section \ref{sec: discuss1}). Furthermore, we argue that neither of those two techniques should be regarded as more reliable than the other (Section \ref{sec: discuss2}).

\subsection{Is the comparison actually self-consistent?}\label{sec: discuss1}

Our comparison of integrated light versus resolved stellar population studies has the main advantage that it uses the same dataset. Nevertheless, we still need to consider two aspects related to the fitting of the stellar populations, that are both nontrivial to implement: \par
First, we would need to make sure to use the exact same stellar synthesis models. Only then, we would be able to estimate, if discrepancies in the derived age and metallicity properties between the two techniques are induced by the different models or the different approaches themselves. The resolved study from \citet{mayte} uses the ELODIE 3.2 \citep{elodie,elodie3,elodie32} stellar library to estimate the stellar parameters for $4750-6800$ \AA\ and the Dartmouth \citep{dartmouth} isochrones in order to determine the stellar ages. On the other hand, we have used the E-MILES SSP library \citep{vazdekis16} together with the BaSTI isochrones \citep{basti04} to fit for ages and metallicities in the wavelength range of $4750-8950.4$ \AA. In this wavelength regime, the E-MILES SSP models are based on three different stellar libraries: the MILES \citep{miles}, the near-IR CaT \citep{cat} and the Indo-U.S. \citep{indous} library \citep[see also][]{vazdekis12}. With these differences basically all systematics regarding spectral and stellar synthesis as well as their respective modelling are captured. Different sets of isochrones have different assumptions about stellar evolution, in- or exclude certain evolutionary phases of stars and are computed for different age and metallicity bins. Different stellar libraries have different flux calibration, wavelength coverage and, in case of empirical ones, are biased towards metallicities in the solar neighbourhood. All of this can influence the derived \emph{absolute} ages and metallicities of stellar populations in \emph{both} analysis techniques neglecting additional systematics induced by, for example, individual element abundances \citep{vazdekis01,schiavon02} as discussed in Section \ref{sec: results}. \par
In principle, this issue could be resolved by using the exact same models in both approaches. Where the authors of the resolved analysis have in theory the freedom to choose any combination of stellar libraries and set of isochrones, analysis of integrated light is restricted to the publicly available SSP models, which have a fixed combination of stellar library and isochrones\footnote{However, tools like FSPS \citep{conroy09,conroy10b} try to provide users a more flexible interface in calculating SSP models.}. Changing models and re-doing the integrated analysis is straightforward, but in the resolved case steps $2.-5.$ in Section \ref{sec: mayte_analysis} have to be repeated, which can be quite time consuming for a decent amount of models and are out of the scope of this study. \par
For the interested reader, we show in Appendix \ref{appendix2} our integrated analysis of M\,54 conducted with the PEGASE-HR SSP library \citep{pegasehr}, in order to match the stellar library used in resolved method, however there the isochrones are from PADOVA \citep{padova94}. It is quite interesting to see how much the recovered age-metallicity distribution seems to depend on the adopted SSP models at first glance, while they are still recovering the same physical implications of M\,54's multiple populations, all modelling uncertainties considered.\par
The second aspect that arises when trying to compare the two approaches is that the color code in Figure \ref{fig: comparison} does not represent the same physical quantity. The resolved study shows the number of stars in each age-metallicity bin, but the integrated analysis provides us with a mass (or light) fraction of an SSP corresponding to a particular age and metallicity. From the latter, we can deduce the absolute mass in each bin and consequently the total mass of M\,54 relatively easily. Inferring the mass of each star can be deduced from the fitted isochrone and their magnitudes \citep[see e.g. also][]{pont04,lin18}. However, a completeness correction is necessary to account for non-detected stars below the turn-off, where most of the mass lies. Only then could the stellar mass in each age-metallicity bin for the individual star analysis be estimated. The individual star counts also influence the derived mean ages and metallicities making the comparison of their values to the integrated measurements not one-to-one let alone in a light- or mass-weighted sense.

\subsection{Is one method more reliable than the other?}\label{sec: discuss2}

After discussing the potential ways of making the comparison between the resolved and integrated analysis of M\,54 as self-consistent as possible, the question arises, whether we would gain new knowledge from this extra work. Certainly, this would put the two approaches to the ultimate test, but we would also need to assume that one approach is better or more reliable than the other. In fact, both approaches suffer from the same difficulties that are connected to the well-known degeneracies and difficulties in stellar population modelling, such as the age-metallicity degeneracy, unknown individual element abundances and poor age resolution at old ages ($\gtrsim 8$ Gyr). \par
With regard to the resolved stars from the MUSE data, their iron abundance is generally well-determined by their spectra, however the age determination from isochrone fitting is rather degenerate, as especially other element abundances shift the isochrones and can induce artificial age variations (see Section \ref{sec: results}). Furthermore, fitting an isochrone through one point on the CMD is very degenerate in itself, as can be seen from the outliers in Figure \ref{fig: comparison} a). Some kind of measure needs to be defined in order to identify these outliers, as was done in \citet{mayte} with the means of Gaussian mixture models.\par 
Similarly, in the integrated analysis, SSP models at old ages ($\gtrsim 8$ Gyr) and at fixed metallicity are more or less indistinguishable, therefore the age leverage is poor in this regime. This essentially could mean that the two old metal-poor populations in Figure \ref{fig: comparison} b) (one at around 8 Gyr and the other one at 14 Gyr) are the same, which is further complicated by the high oxygen abundances in M\,54 as discussed in Section \ref{sec: results}. Furthermore, here it is also hard to tell, which mass weights are robust and which are erroneously being generated simply due to the ill-posed nature of the inversion problem. Nevertheless, the metallicity determination seems to be more robust and hence having a handle on the metallicity distributions from integrated spectra of extragalactic objects is already a big advantage as compared to average values.\par
A similar argument holds when it comes to the comparison of different SSP models with the same technique. We do not know which models are intrinsically more reliable or closer to truth in nature, although a good parameter coverage across $\log \mathrm{g}$, $\mathrm{T}_{\mathrm{eff}}$ and [Fe/H] (and potentially [$\alpha$/Fe]) of stellar spectra is always a limiting factor. \emph{A} choice of a certain SSP library always has to be made, therefore absolute quantities, such as the exact position of the mass weights in age-metallicity space might not be as reliable. However, the relative trends are expected to be the same. Meaning that, if we always use the same SSP models for different stellar objects of interest, we will be able to say differentially, whether the objects have experienced different chemical enrichment histories. \par
In conclusion, without being prejudiced against the credibility of either of the two methods, we can reliably say that \emph{both} results in Figure \ref{fig: comparison} show the following results:
\begin{enumerate}
    \item There are \emph{multiple} stellar populations present in M\,54.
    \item Overall, there is a division between an old, metal-poor ($>8$ Gyr and $\sim$ -1.5 dex) and a young, metal-rich ($1-2$ Gyr and $\sim 0.00-0.25$ dex) population.
    \item The cluster is dominated by the old, metal-poor population.
    \item The mean age and metallicity are in the range of $9-9.5$ Gyr and -1.0 to -1.2 dex.
\end{enumerate}
Hence, the integrated analysis is capable of identifying multiple stellar populations from a single integrated spectrum. It results in a similar star formation and enrichment history as the resolved analysis based on CMD analysis and spectral fitting of individual stars.

\section{Conclusions}\label{sec: conclusion}

In this work we have presented the analysis of M\,54's integrated spectrum from three different data sets (MUSE WFM, MUSE WFM-AO and WiFes) with the goal to recover its multiple stellar population content via full spectral fitting \citep[pPXF:][]{ppxf04,ppxf17} of a library of SSP models \citep[E-MILES:][]{vazdekis16} to the observed spectrum. Thanks to the individually extracted stellar spectra of the $3.5^{\prime}\times 3.5^{\prime}$ MUSE WFM data set from \citet{mayte}, we could also investigate the influence on the stellar population recovery by excluding the contribution of certain stars to the total integrated spectrum. In light of all our tests, we draw the following conclusions in recovering age-metallicity distributions from integrated spectra:
\begin{itemize}
    \item The derived mass fractions in age-metallicity space are robust against 1) the Na notch filter in MUSE-AO observations (Figure \ref{fig: all_cubes}), 2) the inclusion of stars classified as non-members (Figure \ref{fig: single_stars}) and 3) the contribution of extended horizontal branch stars (Figure \ref{fig: horizontal_branch}).
    \item The recovery of the age-metallicity distribution is not very sensitive to the limiting magnitude of the observations. Consistent results are achieved, even if the limiting magnitude were 4 times brighter than the main sequence turn-off region (Figure \ref{fig: magnitude_cut}).
    \item The recovered mass fractions are consistent in their absolute position in age-metallicity space over individual pointings of the $4\times4$ MUSE mosaic, as long as the spectrum of an overly bright star does not dominate the integrated spectrum (Figure \ref{fig: single_cubes}).
    \item Additional spectral coverage in the bluer wavelength ($3500-4000$ \AA) does not change the age-metallicity distribution recovery significantly (Figure \ref{fig: waggs1}, \ref{fig: waggs2}).
    \item Bright ($\mathrm{I<14}$ mag) and red ($\mathrm{V-I}>1.7$ mag) stars in the integrated spectrum seem to induce erroneous old ($>8$ Gyr), metal-rich (+0.25 dex) populations in the recovered age-metallicity distribution (Figure \ref{fig: single_stars}). Uncertain evolutionary phases such as the thermally pulsing AGB not included in the SSP models could be an explanation for this.
    \item The absolute derived ages and metallicities change, as expected, with different SSP model assumptions, however differentially the trends stay the same (Figure \ref{fig: comparison}, \ref{fig: pegasehr}). Hence, M\,54's \emph{multiple} stellar populations are indeed retrievable from its integrated spectrum showing an old ($8-14$ Gyr), metal-poor (-1.5 dex) as well as a young (1 Gyr) and metal-rich (+0.25 dex) population.
    \item The derived mass-weighted mean age and metallicity of 9.53 Gyr and -1.21 dex are consistent with the corresponding averages of the resolved analysis of 9.24 Gyr and -1.01 dex respectively.
    \item The derived stellar M/L ratios show more stochasticity in the outer regions of M\,54 (M/L$_{\mathrm{V}}=1.3-1.8$), where the luminosity density is lower, as compared to the central region, where the value converges to around 1.46. We attribute this to the lower relative contribution of young, metal-rich mass fractions.
\end{itemize}
In this context we also compared and discussed our results with findings of the resolved stellar population analysis from same MUSE WFM data set \citep{mayte}. From this we find that age-metallicity \emph{distributions} can be derived from full spectral fitting of integrated spectra with comparable reliability as from resolved studies, as both approaches suffer equally from the same difficulties, uncertainties and degeneracies in stellar population synthesis modelling, especially with regard to age determinations, whereas the recovered metallicity distribution seems to be more robust. While IFU observations of resolved systems can certainly provide detailed information on a star-by-star basis, our integrated approach can provide the same information content, if the scientific goal is to disentangle multiple or complex stellar populations of stellar systems. It is also worth noting that the integrated analysis reveals this information with a single fit in several minutes as opposed to lengthy data extraction and analysis steps undertaken in the case of the resolved study (see Section \ref{sec: mayte_analysis}). On top of that, our returned distributions have a physical unit attached to them (mass fractions) instead of number counts, which lets us straightforwardly calculate the stellar mass of the different populations or the system in total as well as M/L ratios. This provides us with a quick and detailed knowledge about the stellar content of an object. \par
In spite of the modelling differences between both methods, we find that the average age and metallicity from the integrated and resolved stars analysis agree remarkably well with each other. This is of key importance for extragalactic studies at low and high redshift, which can only access the integrated light - especially now with the advanced development of chemo-dynamical models for external galaxies \citep[e.g.][]{poci19,zhu20}. \par
With the ability to study the two dimensional distribution in the age-metallicity plane of thousands of stellar objects, we can establish a connection between the properties of multiple stellar populations to their global properties like total mass, presence of super-massive black holes and environment. This means, we are now in an era (data- and modelling-wise) to constrain formation scenarios of nuclear star clusters or the stellar mass assembly of galaxies on a statistical scale with the stellar population distributions from integrated spectra.

\acknowledgments

We would like to thank the anonymous referee, who provided a very careful and constructive revision that improved this manuscript. AB likes to thank Iskren Georgiev and Nikolay Kacharov for their help in this project. We also thank Anil Seth for useful discussions. This work was funded by the Deutsche Forschungsgemeinschaft (DFG, German Research Foundation) -- Project-ID 138713538 -- SFB 881 (``The Milky Way System'', subproject B08). IMN acknowledges support from the AYA2016-77237-C3-1-P grant from the Spanish Ministry of Economy and Competitiveness (MINECO) and from the Marie Sk\l odowska-Curie Individual \textit{SPanD} Fellowship 702607. RL also acknowledges support from the Natural Sciences and Engineering Research Council of Canada PDF award and DAAD PPP exchange program Projekt-ID 57316058. This research has made use of the services of the ESO Science Archive Facility.

%






\bibliography{m54}
\bibliographystyle{aasjournal}

\appendix

\section{Quantitative comparison between recovered mass fractions}\label{appendix3}

\begin{figure*}[htbp]
\centering
\includegraphics[width=\textwidth]{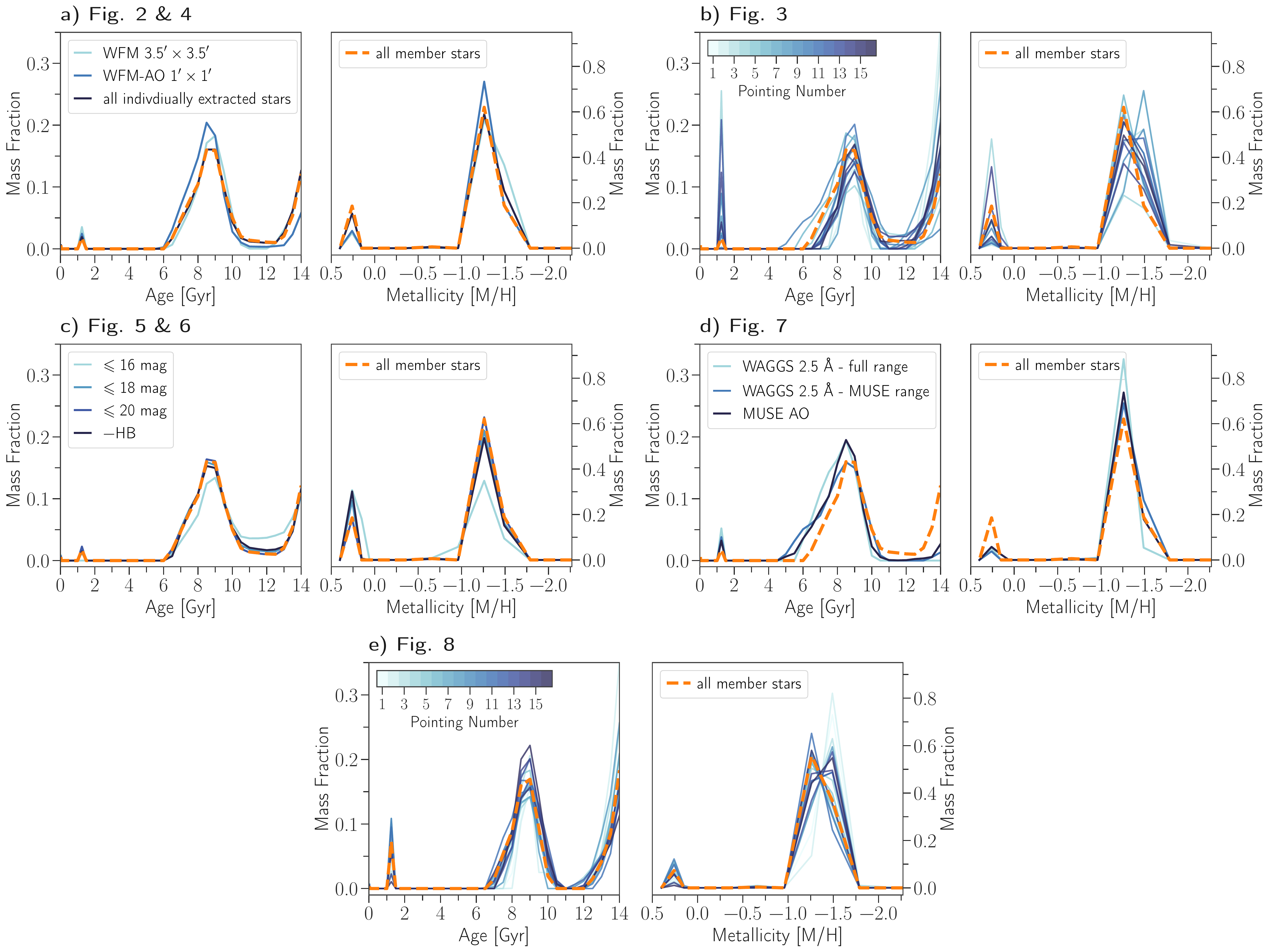}
\caption{\textit{a):} One dimensional distributions of mass fractions as a function of age and metallicity respectively. They are derived from the integrated spectra presented in Figures \protect\ref{fig: all_cubes} \& \protect\ref{fig: single_stars}. The orange dashed line shows the mass distribution for the integrated spectrum made from individual member stars. It is shown in all panels in order to ease comparison between the different results from the integrated spectra investigated in Section \protect\ref{sec: tests1}. \textit{b):} The same for Figure \protect\ref{fig: single_cubes}. The color-code shows the corresponding pointing number as in Figure \protect\ref{fig: single_cubes}. \textit{c):} The same for Figures \protect\ref{fig: magnitude_cut} \& \protect\ref{fig: horizontal_branch}. \textit{d):} The same for Figure \protect\ref{fig: waggs1}. \textit{e):} The same for Figure \protect\ref{fig: stars_per_cube}. The orange dashed line shows now the results for all member stars without the contribution of the brightest stars as defined in Section \protect\ref{sec: comp}.}
\label{fig: summary}
\end{figure*}

We show in Figure \ref{fig: summary} the results from fitting the different integrated spectra of M\,54 from Section \ref{sec: tests1} in the form of one dimensional distributions as a function of age and metallicity respectively. This allows for a more quantitative comparison between the different fits than the two dimensional age-metallicity distributions. Evidently the recovered mass fractions show overall consistent results between the various integrated spectra that we investigated. Comparing panels b) and e) in Figure \ref{fig: summary} we see that the derived mass fractions across all 16 independent MUSE pointings show less variations, especially in the young (1 Gyr), metal-rich (+0.25 dex) component, when foreground stars and the brightest member stars are not included in the resulting integrated spectrum.

\section{Color-magnitude diagrams of M\,54 in the 16 MUSE pointings}\label{appendix4}

\begin{figure*}[htbp]
\centering
\includegraphics[width=0.75\textwidth]{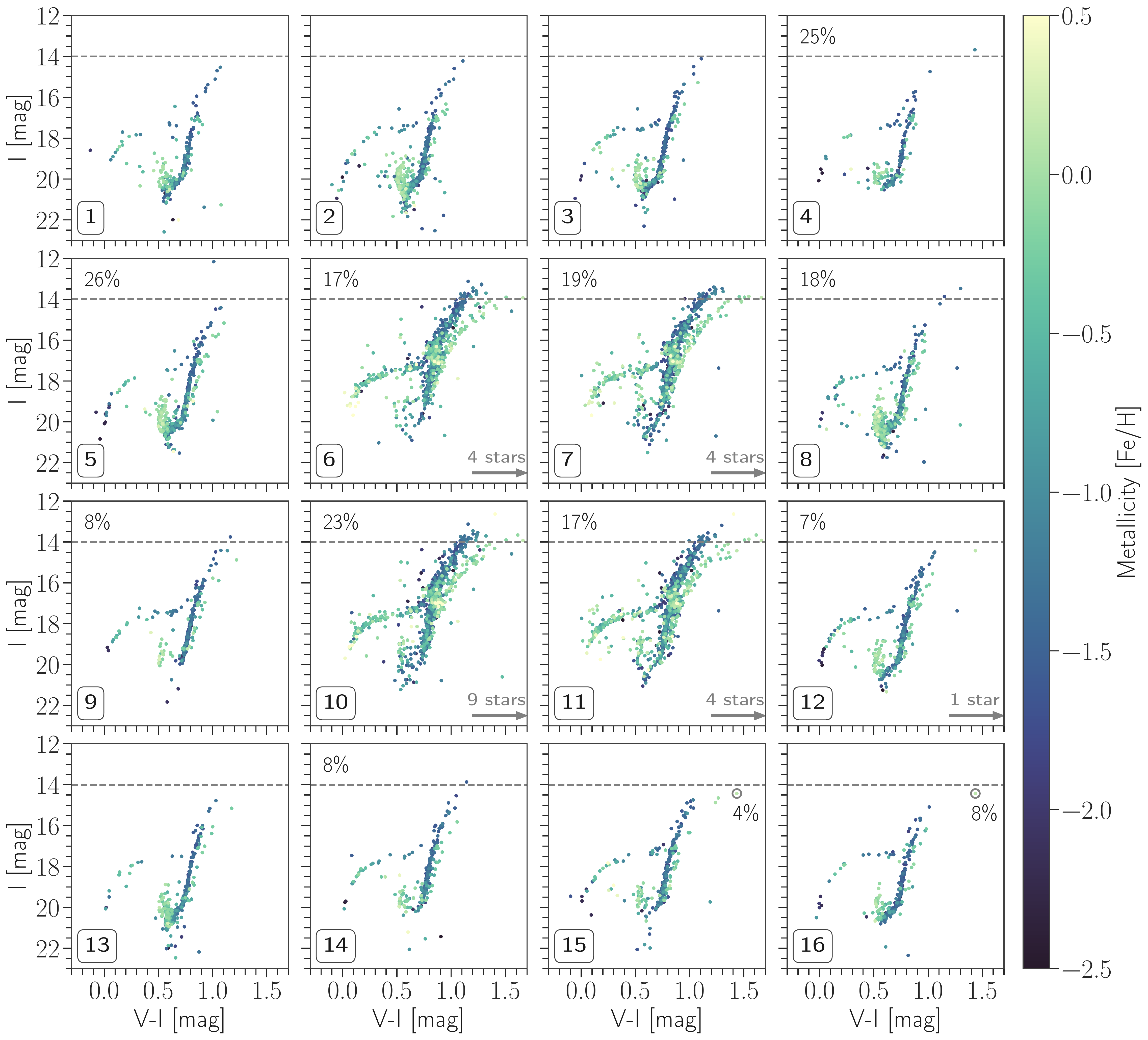}
\caption{Color-magnitude diagram for members stars of M\,54 that where extracted from the corresponding MUSE pointing. The color code follows the measured metallicity [Fe/H] for each star from \protect\citet{mayte}. They are arranged in the same order as they appear on the sky in Figure \protect\ref{fig: data}. The grey arrow indicates how many stars have extremely red colors ($\mathrm{V-I}\gg1.7$ mag) and are thus not shown. The percentage in the upper left hand corner indicates the fraction of flux of stars that are brighter than 14 mag in I and redder than 1.7 mag in $\mathrm{V-I}$. In pointing number 15 and 16 the percentage shows the flux contribution of the same star (grey circle) to total flux of member stars in that pointing.}
\label{fig: cmd_per_cube}
\end{figure*}

Figure \ref{fig: cmd_per_cube} shows the CMDs of M\,54 member stars extracted from their corresponding MUSE pointings from the $3.5^{\prime}\times 3.5^{\prime}$ mosaic \citep{mayte}. Especially, the central pointings (number 6, 7, 10 and 11) have a high number of bright ($\mathrm{I<14}$ mag) and extremely red ($\mathrm{V-I}\gg1.7$ mag) stars that contribute around 20\% to the total flux of the integrated spectrum made from member stars in that corresponding MUSE pointing. When these stars are excluded from the integrated spectrum, the contribution of the previously recovered old ($>8$ Gyr) and metal-rich (+0.25 dex) mass fractions decreases from $20-30\%$ to under 2\%. In pointing number 16 a single star with 8\% flux contribution was responsible for pPXF to recover these mass weights. The same star is also present in pointing number 15, where it did not cause the old, metal-rich component to be picked up, possibly because its flux contribution is decreased to only 4\%. Interestingly, other pointings (e.g. 4, 5 and 8) also have a few bright stars that contribute a significant amount to the total flux, but did not cause any old, metal-rich component to appear in the derived age-metallicity distribution. A possible explanation might be that these stars (expect the bright star in pointing 4) do not have red continuum shapes as well as TiO absorption bands that become visible in the integrated spectrum, which could easily mimic old, metal-rich stellar populations in the SSP models.

\section{Fitting the MUSE integrated spectrum beyond 8950.4 \AA\ with the E-MILES library}\label{appendix1}

\begin{figure*}[htbp]
\centering
\includegraphics[width=\textwidth]{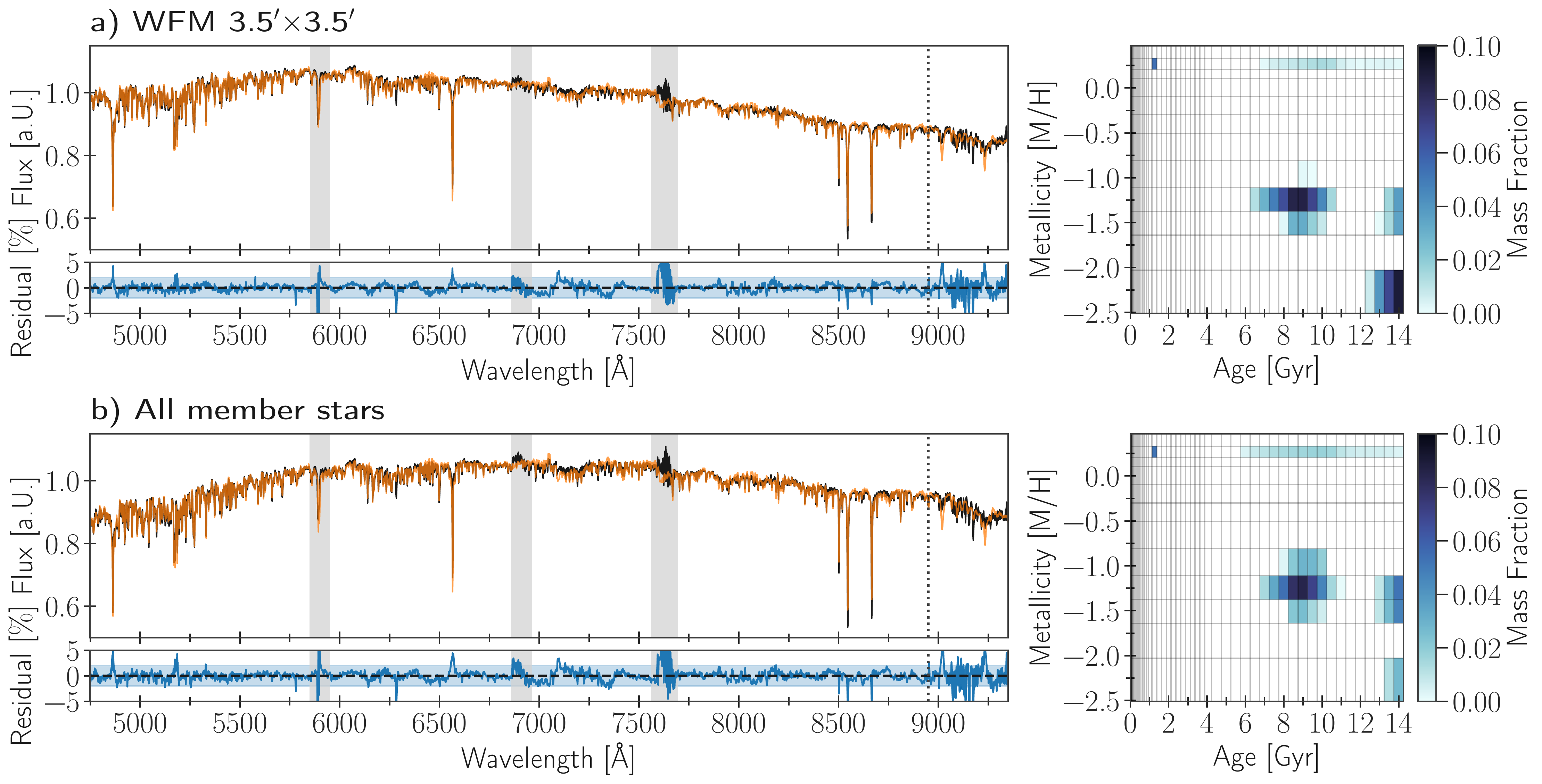}
\caption{\textit{a):} The left panel shows the pPXF fit (orange) to the integrated spectrum of M\,54 (black) from the combined $3.5^{\prime}\times 3.5^{\prime}$ MUSE WFM mosaic now fitting the \emph{entire} MUSE wavelength range. The residuals are shown in blue and the corresponding band shows the 2\% level. Grey shaded areas are masked out sky residuals. The grey dotted lines shows the previous wavelength cut-off at 8950.4 \AA. The right panel shows the derived mass fractions in age-metallicity space that make up the bestfit from pPXF. \textit{b):} The same as for a) but now showing the integrated spectrum from individual member star spectra.}
\label{fig: EMILES_cut}
\end{figure*}

As mentioned in Section \ref{sec: SSP} we here present the mass distribution in age-metallicity space derived from the integrated spectrum of the entire $3.5^{\prime}\times 3.5^{\prime}$ MUSE mosaic and made from the individually extracted member stars \citep{mayte} fitted to the entire wavelength range of MUSE with the E-MILES models. The results are shown in Figure \ref{fig: EMILES_cut}. We see that pPXF now recovers a third component, which is at 14 Gyr and the lowest metallicity of -2.27 dex. It is very prominent, when the integrated spectrum was made from all 16 MUSE data cubes combined. Apparently, the recovery of this component depends on the wavelength range between $8950.4-9300$ \AA, which is quite noisy, as can be seen from the residuals. Moreover, in this wavelength regime the E-MILES models have a lower resolution ($\mathrm{FWHM}\approx4.2$ \AA) than the MUSE spectrum, which makes the two Paschen lines ($\mathrm{n}=9$ and $\mathrm{n}=10$) appear very broad. Nevertheless, in the observed spectrum they do not appear nearly as deep as the bestfit model. \par
Following the discussion in Section \ref{sec: results} we argue that this does not have any influence on our statements regarding the ability and reliability of recovering multiple populations from integrated spectra, as the exact absolute values of the ages and metallicites depend on the adopted SSP models. We can still make the same qualitative statement about M\,54's multiple stellar populations and now the overall trend of the recovered chemical enrichment is even more consistent with a steady rise than when the wavelength rage was cut off. \par
To assess whether the broader spectral resolution of the E-MILES in that particular wavelength range could cause the new stellar population component, we have convolved the integrated spectrum to the lowest spectral resolution present in the E-MILES library at those wavelengths (FWHM 4.4 \AA). Still the 14 Gyr old component in the lowest metallicity bin was recovered.

\section{Fits with the PEGASE-HR SSP library}\label{appendix2}

\begin{figure*}[htbp]
\centering
\includegraphics[width=0.75\textwidth]{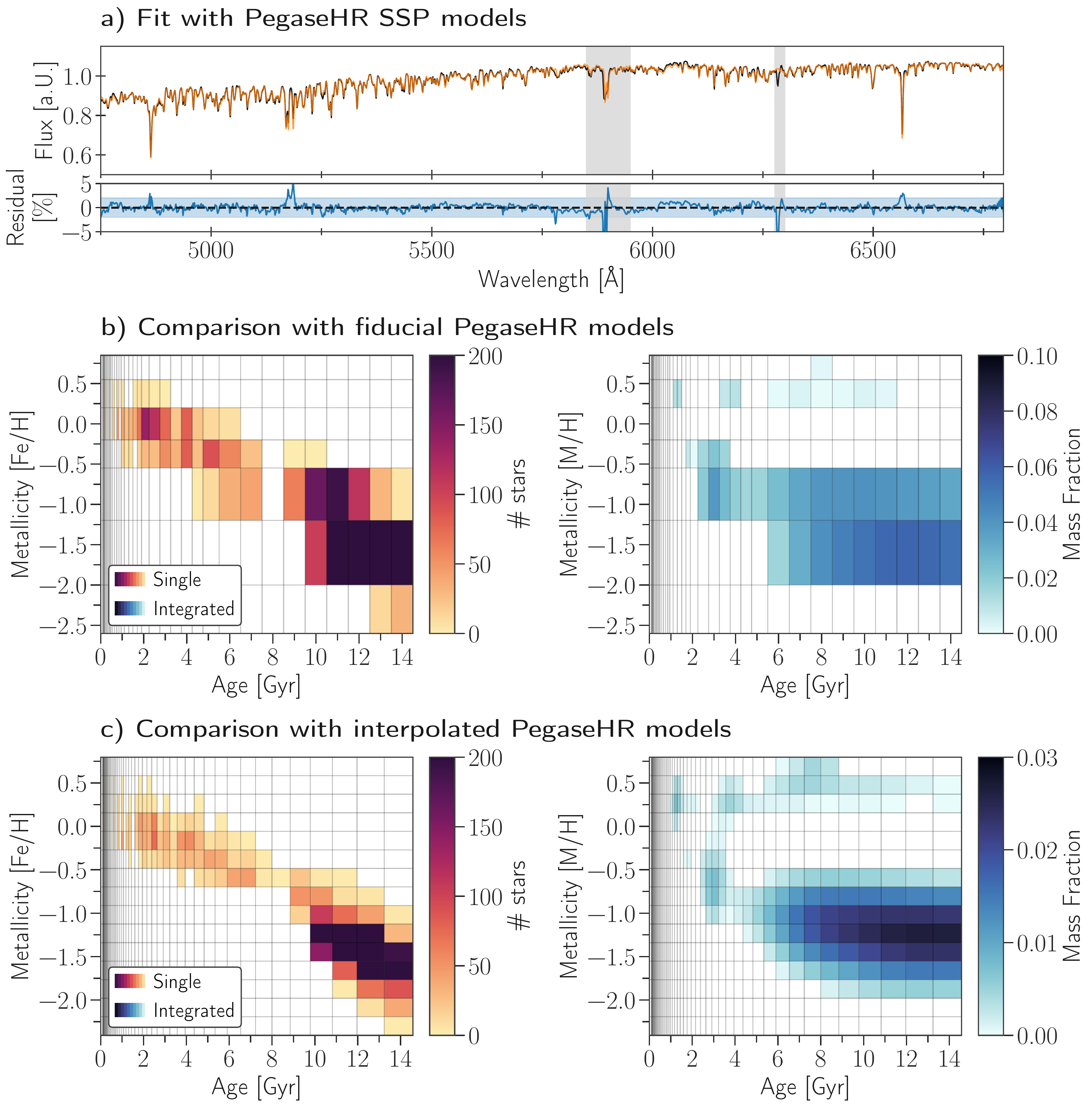}
\caption{\textit{a):} pPXF fit (orange) to the integrated spectrum of M\,54 (black) constructed from all individually extracted member stars \protect\citep[see][]{mayte} with the PEGASE-HR SSP models. The residuals are shown in blue and the corresponding band shows the 2\% level. Grey shaded areas are masked out regions. \textit{b):} Comparison between the resolved (left) and integrated (right) results of M\,54's stellar populations. The integrated analysis has been conducted with the fiducial PEGASE-HR models. \textit{c):} The same comparison as b), but here the PegaseHR models where interpolated onto a finer age-metallicity grid prior to fitting. Here, the individual mass weights returned by pPXF are on average much lower than with the fiducial model.}
\label{fig: pegasehr}
\end{figure*}

Here, we provide results of fitting the integrated spectrum of M\,54 with the PEGASE-HR models \citep{pegasehr}. They are based on the ELODIE 3.1 \citep{elodie,elodie3,elodie31} high resolution spectra ($R=10000$) and the Padova isochrones \citep{padova94}. The wavelength coverage is $3900-6800$ \AA, ages and metallicity span $0.001-20$ Gyr (68 bins) and -2.3 to 0.69 dex (7 bins) respectively. The assumed IMF is Kroupa \citep{kroupa01} and the mass of one SSP is also normalized to unity. We set the minimum age bin to 0.1 Gyr and the maximum to 14 Gyr in order to match the boundaries of the E-MILES models, however we made sure that no spurious mass weights were detected when including the full age range in PEGASE-HR models. Similarly to \citet{kacharov18}, we also performed a pPXF fit with PEGASE-HR models that were interpolated to a finer age-metallicity grid. The models were fit to the integrated spectrum made from the individually extracted member star spectra. Both results and the comparison to the resolved study from \citet{mayte} are shown in Figure \ref{fig: pegasehr}. \par
We again detect an old, metal-poor population with a metallicity between -1.0 and -1.5 dex and an essentially unconstrained age between 8 and 14 Gyr. Now, we can also identify an intermediate population at around 3 Gyr and between -1.0 and -0.5 dex in metallicity, which is still more metal-poor than the intermediate-age population from \citet{mayte}. We also still see the young, metal-rich population at 1 Gyr and around +0.25 dex, again more metal-rich than the one identified in the resolved analysis. Mass weights with this same super solar metallicity but older ages are again attributed to the same systematics as discussed in Section \ref{sec: results}. \par
The result from the interpolated PEGASE-HR models shows mass weights that are in the same location in the age-metallicity space as the fiducial models, but are on average lower. The weights are hence smeared out across several more age-metallicity bins, which is not surprising as the interpolation adds more linearly dependent models into the design matrix. \par
\begin{figure*}[htbp]
\centering
\includegraphics[width=\textwidth]{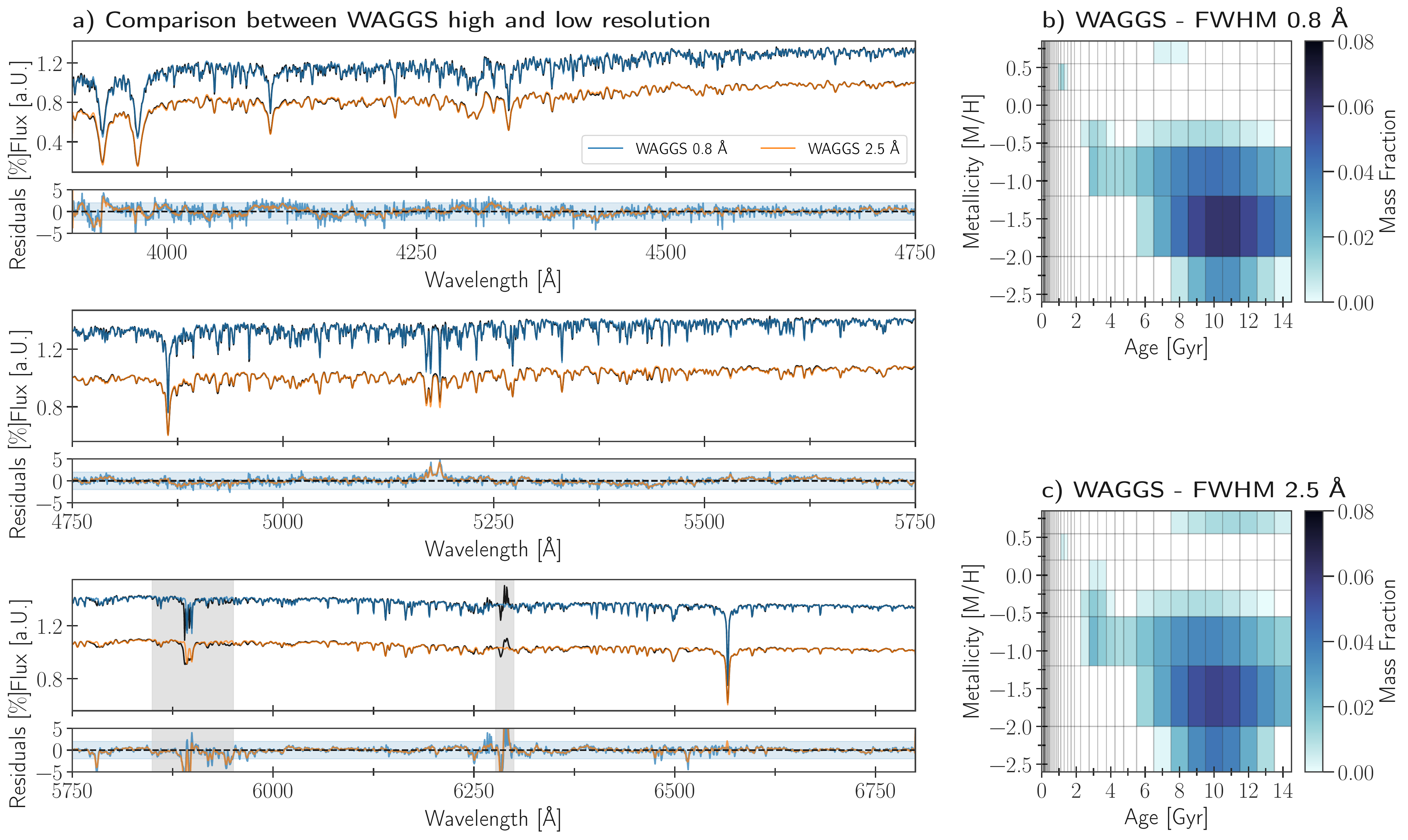}
\caption{\textit{a):} pPXF fit to the integrated spectrum of M\,54 from WAGGS \protect\citep{waggs1} with the PEGASE-HR models. The fit to the native spectral resolution (FWHM 0.8 \AA) is shown in blue, whereas the broadened one (FWHM 2.5 \AA) is in orange. The residuals are on the 2\% level for both resolutions. \textit{b):} Recovered age-metallicity distribution from the high resolution spectrum. \textit{c):} Same as b), but for the broadened spectrum with a FWHM comparable to MUSE.}
\label{fig: waggs2}
\end{figure*}

In Figure \ref{fig: waggs2} we show the results of fitting the WAGGS spectrum of M\,54 with the fiducial PEGSASE-HR SSP models, once for the native WAGGS spectral resolution (FWHM 0.8 \AA) and once broadened to 2.5 \AA\ to mimic the spectral resolution of MUSE. Both results are almost identical, whereas the high resolution fit retrieves much less mass weights at old age and super solar metallicity and gives more weight to the old, metal-rich (around 10 Gyr and -1.5 dex) and the young, metal-rich (around 1 Gyr and +0.25 dex) population as compared to the result from the lower resolution spectrum. We can also see some differences in the distribution of the recovered mass fractions in age-metallicity space, if we compare these to the results for the MUSE spectrum fitted with the PEGASE-HR models in Figure \ref{fig: pegasehr}. Nevertheless, this gives us confidence that the recovery of ages and metallicities of multiple stellar populations from integrated spectra is not severely dependent on the spectral resolution. \par
Even though the fit to the WAGGS spectrum with the PEGASE-HR models in Figure \ref{fig: waggs2} includes bluer wavelengths than the MUSE data, we still recover an extended old, metal-poor population, where compared to fitting the full WAGGS wavelength range with the E-MILES models in Figure \ref{fig: waggs1}, the metal-poor mass fractions at 14 Gyr vanished and instead concentrated around 8 Gyr. We argue that this difference arises because of the diverse modelling assumptions of stellar population synthesis and not because of the inclusion of bluer wavelengths.

\end{document}